# Typologies of the Popular Science Web Video

Jesús Muñoz Morcillo, Klemens Czurda, Caroline Y. Robertson-von Trotha
ZAK | Centre for Cultural and General Studies at the Karlsruhe Institute of Technology (KIT)

## 1. Abstract

The creation of popular science web videos on the Internet has increased in recent years. The diversity of formats, genres, and producers makes it difficult to formulate a universal definition of science web videos since not every producer considers him- or herself to be a science communicator in an institutional sense, and professionalism and success on video platforms such as YouTube no longer depend exclusively on technical excellence or production costs. Entertainment, content quality, and authenticity have become the keys to community building and success. The democratization of science video production allows a new variety of genres, styles, and forms. This article provides a first overview of the typologies and characteristics of popular science web videos. To avoid a misleading identification of science web videos with institutionally produced videos, we steer clear of the term "science communication video", since many of the actual producers are not even familiar with the academic discussion on science communication, and since the subject matter does not depend on political or educational strategies. A content analysis of 200 videos from 100 online video channels (190 of them from YouTube, 8 from Vimeo, and 2 from a proprietary vlog) was conducted. Several factors such as narrative strategies, video editing techniques, and design tendencies with regard to cinematography, the number of shots, the kind of montage used, and even the spread use of sound design and special FX point to an increasing professionalism among science communicators independent of institutional or personal commitments: in general, it can be said that "supposed" amateurs are creating the visual language of science video communication. This study represents an important step in understanding the essence of current popular science web videos and provides an evidence-based definition as a helpful cornerstone for further studies on popular science web videos and science communication within this kind of new media.

## 2. Introduction

Since the creation of YouTube in 2005, there have been opposing positions on the quality of web videos. Some criticize the banality displayed by the majority of amateur movies on the Internet (Keen 2007, p. 5; Lovink 2011, p. 9), while others praise the *participatory culture* fostered by this new mass media phenomenon (Jenkins 2006). Today, there is a broad consensus that most videos on the Internet involve familiar or commonplace contents (Marek 2013, p. 17). However, there are some exceptions. One of them is the "popular science web video". A popular science web video is a short video that focuses on the communication of scientific contents for a broad audience on the Internet. For the sake of terminological simplicity and easier reading, in the following we will speak of "science web video" or "science video," referring to the above definition. A set of science videos uploaded by one user constitutes an online video channel. Video channels that make scientific knowledge accessible to the public are the subject of the present study. Up until now, science video channels and the concomitant global phenomenon of web videos have not been subject to analysis. This is all the more surprising as YouTube—the most popular online video platform with more than 4,000 science channels and 100,000 science videos (Yang et al. 2011)[1]—presents itself as a highly visible, varied, and growing data corpus with worldwide accessibility.

From the viewpoint of science and technology studies, which examine inter alia the impact of new media on science communication (cf. Bucchi 1998, 2008; Robertson-von Trotha 2012), some key questions arise: What are the main characteristics of popular science web videos? Who is communicating science through these videos, and for what purpose? And how are these videos related to the main characteristics of the overall phenomenon of video communication on the Internet?

The following data analysis of 200 web videos provides a general typology of the global operating tendencies of science and educational video channels on the Internet. This is meant to constitute a first step for further investigations on this particular form of the web video, its production context, and its importance for science communication. The main goals of our analysis are: (1) the identification of the most popular science video channels and their producers according to the findings of the YouTube search algorithm both worldwide and in each individual country;[2] as an additional criterion, we compare these results with the recommendations of experts on reputable online science blogs; (2) producing a typological study on aesthetic and narrative trends on science web videos for the public; and (3) providing an informational basis for future context and network analysis with a focus on the interaction and influence between science videos and their creators on the Internet.

This introduction is followed by an outline of the methodology used, in which the data corpus is defined. Furthermore, we present a description of the quantitative data analysis carried out by means of a standardized codebook. Finally, we present our findings and the conclusions drawn from them, providing an outlook on possible future research.

# 3. Methodology

### The Organizational Structure of YouTube

To form a methodology analyzing popular scientific web videos, it is first necessary to sketch an outline of YouTube and how it organizes its online appearance. At present, YouTube is the most popular platform for web videos (Alexa 2015; Quantcast 2015). We sampled 190 videos related to science and education published on the Google-owned website, but we also included, as references, 10 videos from two other websites: the video platform Vimeo and the university project Teslablog.

On the front page as well as on the subpages of YouTube, videos are listed as so-called "thumbnails", small images that show a still from the video in question. YouTube displays its content in 77 different localizations[3] and 61 languages based on the accessing users' IP addresses and browser settings. The localized pages are organized in different thematic sections, called lists. These lists basically consist of channels with uploaded videos. The channels are maintained by user accounts belonging to individuals, groups, companies, or other governmental and non-governmental agencies generally called "YouTubers". As for the video site: A video is separated into the video itself, offering embedded text and links, and the website framing it contains an information area and a comment section with sharing options as well as a voting system consisting of "thumbs up" and

[1] Every minute, 300 hours of video are uploaded on YouTube, according to the company's statistics: https://www.youtube.com/yt/press/statistics.html as of 03.18.15.
2 YouTube is a methodological choice that forms the basis for our data collection.
[3] 1 worldwide setting and 76 country settings: see the bottom section of https://www.youtube.com/, retrieved on 03.18.2015.

“thumbs down” buttons. A channel’s popularity and therefore its appearance on the aforementioned lists is not only determined by the generated views, comments, and “likes”, but also by the number of subscribed users. All these factors influence the popularity of certain videos, which become economic factors too. Accordingly, many YouTubers make a living out of generating YouTube content.[4] For the leading YouTube video creators, providing content is a full-time profession. Since December 2010, the time limit of first 10 and then 15 minutes was completely removed by YouTube, enabling established users to upload videos of any length.[5]

## Selection of Science Web Video Channels

For the selection of science web video channels, we defined the following steps. First of all, it was necessary to search for suitable YouTube channels with disabled cookies and cleaned cache memory data because these factors can interfere with the reliability of the findings due to search personalization settings. For this purpose, we used the “worldwide” list at the YouTube channel category site “Science & Education” (https://www.youtube.com/channels/science_education). This site works with an algorithm that takes not only views and subscription numbers into account, but also user engagement—at least since the end of 2012.[6] This procedure allows the compilation of a global list of popular science channels. The worldwide “Science & Education” list contains roughly the hundred most popular YouTube videos globally, according to YouTube’s own algorithms, which are subject to change. Second, in order to offer a channel selection of science web videos that is as comprehensive as possible, a comparison of channels by country was required. For this purpose, we searched for the most popular science video channels by country. On YouTube it is possible to choose from among the settings of 76 countries (as of 03.18.2015). As a result, roughly one hundred national and foreign science channels that are popular in the selected country are displayed. These results were compared with the previous YouTube global list of most popular science video channels worldwide in order to achieve a reliable YouTube list of the most popular global operating science channels for the general public. Successful national channels (in Spanish, French, Portuguese, and Italian) were included in the sample of YouTube channels studied.

The selection of science web video channels was supplemented by information that we retrieved from highly frequented science blogs. Among these were Open Culture, Getting Smart, Make Use Of, MathsInsider, and others. We identified a total of 63 science blogs by means of Google searches using the following terms: “(best) youtube science channels”, “(best) youtube educational channels”, “science blog youtube”, “recommend(ed) science channels”, and the corresponding translations in Spanish, French, Portuguese, and German. We consulted 31 English, 15 Spanish, 13 German, and 4 Portuguese blogs. Expert recommendations enabled us to triangulate our observations and helped us choose the seemingly more impactful channels, in particular with regard to how often a science video channel was mentioned on the listed blog sites.

To include non-YouTube platforms in the survey, we analyzed 8 videos from vimeo.com and 2 from teslablog.iaa.es. Vimeo is one of the largest video uploading platforms[7]—it is aimed primarily at independent filmmakers and professionals, and offers high-quality content. Teslablog is a specific public science project that promotes an unusual approach by means of web videos and blog entries. This project uses its own dissemination platform, teslablog.iaa.es, and an embedded video player (JWPlayer).

---

[4] In this regard, we analyzed the appearance of subscription requests in the videos or the comment sections of the videos as well as the production background.
[5] See the official YouTube Blog Post of 12.09.2010: http://youtube-global.blogspot.de/2010/12/up-up-and-away-long-videos-for-more.html, retrieved on 03.18.2015.
[6] See http://youtubecreator.blogspot.de/2012/10/youtube-search-now-optimized-for-time.html, retrieved on 03.18.2015.
[7] According to the Internet traffic analyzing companies Alexa’s and Quantcast’s rankings, retrieved on 03.18.2015, http://www.alexa.com/siteinfo/vimeo.com and https://www.quantcast.com/vimeo.com.

### Limitations of the YouTube Search Algorithm and Network Analysis

The YouTube search algorithm changes from time to time without warning or advance notice in academic publications. Because of this, it is reasonable to contrast the findings on YouTube with an independent source. For this purpose, we consulted specialized Internet sites such as blogs and forum sites, all of which are addressed in this paper as the "blogosphere" for the sake of simplicity.

There are limitations to network-based surveys, in particular when they are based on the study of channel recommendations and video answers (which have since been depreciated by YouTube[8]) as indicators of possible aesthetic influences. Channel recommendations are links to specific YouTube channels that are either spoken or appear as clickable text in the video itself or in the web video description. Usually the links direct to a subscription button for the relevant user's channel. For content providers on YouTube, subscriptions by other YouTube users to their channels are vital for their visibility. These self-organized networks primarily provide more visibility to competing channels, and can be taken into account for a description of the function and interaction of science web video networks on YouTube.

In order to detect and make visible not only the main characteristics of the videos but also the possible qualitative influences between them, it was necessary to design the data collection in two different ways: on the one hand, we analyzed the main aesthetic and narrative characteristics that define these videos, such as storytelling complexity and detailed production aspects. On the other hand, we focused on external and non-formal aspects that allow the reconstruction of interaction patterns between the different channels. As an example, we focused on aspects such as chronological factors, the participation of actors (usually scientists) and technical staff (especially at a local or national level) in different channels, cross references, quotes, and so on.[9] Since the main goal of this paper is to develop a typology of the science web video, a thorough social network analysis will be the subject of future studies, although some of the criteria necessary for such an evaluation were already taken into account for data collection (such as the number and placement of links and the production background).

### Web Video Corpus and Exceptions

The decision of whether or not a video on YouTube is categorized as "scientific & educational" is left solely to the uploading user. The uploader-defined categories lead to a loose use of the terms "science" and "education"; we therefore excluded certain channels from the corpus following the guidelines outlined here.

For the present research we defined "science web video" as an "edited educational or science web video that addresses both young people and adults" and tackles topics related to science, guided by the OECD classification of Fields of Science and Technology (FOS).[10] As a result of this definition, the following channel types were not taken into account: channels with unedited live recorded videos (such as lectures as seen on the MIT-Channel or India NPTEL-Channel as well as MOOCs (Massive Open Online Courses); channels with no more than five of their own uploads; instructional videos (such as tutorials or videos on cosmetic tips); science movies for children; channels containing intriguing information not related to a scientific topic; channels with educational songs; TV science channels (unless they produce original content for the Internet, such as the Discovery Channel or BBC Earth); channels with informational videos on sexuality; channels with spiritual content; channels without their own uploader-generated content; and cooking channels.

---

[8] As of September 2013, the "video response" feature is no longer available on YouTube: http://youtubecreator.blogspot.de/2013/08/so-long-video-responsesnext-up-better.html, retrieved on 03.18.2015.
[9] For more on network analysis applied to film and new media, please consult: Nicholas Christakis and James H. Fowler, 2010, and "The Oracle of Kevin Bacon", Watts and Strogatz, 1998; and for more on network analysis as it is applied to music, please consult Uzzi and Spiro, 2005.
[10] For further reference, see the Frascati Manual of the OECD here: http://www.oecd.org/science/inno/38235147.pdf, retrieved on 03.17.2015.

The resulting video channel list includes 100 representative video channels (see **Table 1**). From every channel we chose the most recent and the most popular video for analysis. In this regard, our study on the typologies of the popular science web video is based on a corpus of 200 web videos (190 from YouTube, eight from Vimeo, and two from Teslablog).

For our study we divided data collection in two phases:

1. Collecting general information about the channel: name, link, video account, subscribers, date of account registration, most popular video (name, link, date, views, likes, dislikes, subtitles), and the last uploaded video (name, link, date, views, likes, dislikes, subtitles).
2. Collecting specific data about the particular videos (the most popular and the most recent videos were chosen): aesthetics and narrative characteristics with standardized form sheets; number of actors according to gender; thumbnail (the video preview still) description; estimated age of the actor(s); shooting location; camera work; average number of shots; kind of storyline; genre as well as intro and outro description; special effects (FX) used; light and sound design; type of music; audio quality; and the quality of the narrator's voice.

In addition, we gathered information about the uploading account such as the user name and the number of views and uploaded videos since the date of registration. We also documented the kind of recommended links, their position on the screen and/or in the description underneath the video, and if these links were additionally spoken by the user or not. Lastly, we posed questions about the production context, e.g. if the video was an individual's work or made by a group of people, and whether or not the video was for profit (which depends on the presence of advertisements before the video). For detailed information about the coding questionnaire, see **Appendix Table 1**.

# 4. Data analysis

### Average Popularity of Science Web Videos

For reasons of statistical consistency, videos from Vimeo and proprietary platforms were not taken into account when calculating average popularity. In order to calculate the average popularity of the selected videos, we divided the number of views by the days a video was online since its publication on YouTube. For this purpose, the newest video and the most popular video of every selected channel were taken into account. It should be noted here that YouTube videos generate a lot of views in the first few weeks after their publication, and that the growth in views decelerates the longer the video is online. Therefore, this statistic favors newer uploads.

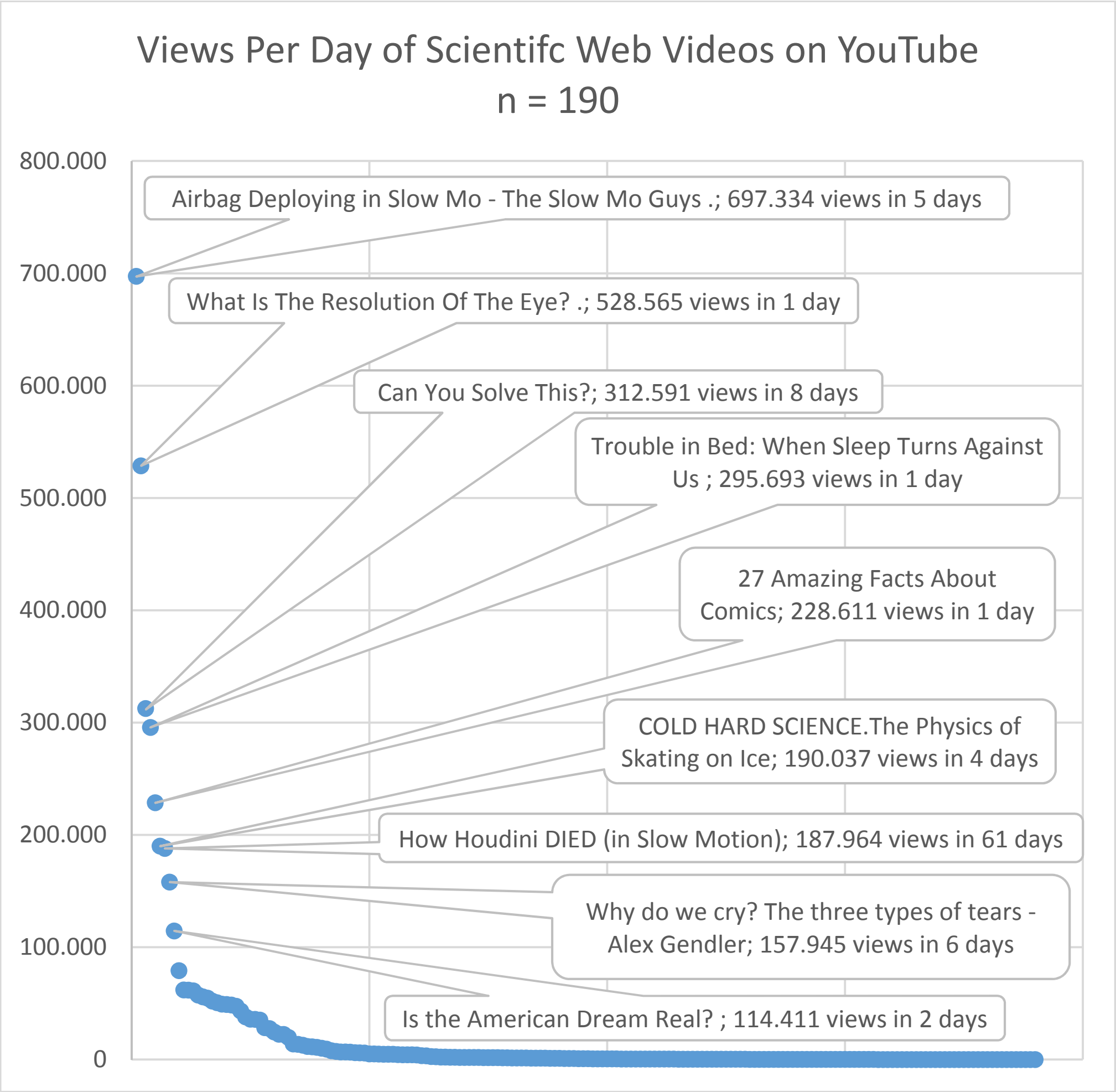

**Figure 1:** Video popularity by means of views per day:
a few spikes with more than ½ million and ¼ million views per day.

Although we consider all selected videos popular enough to be listed, our research also examines the quantitative differences between them, in order to find what kind of content and format is most popular, even if this method only serves as a first approach.

Most of the analyzed videos have between 100 and 6,500 views per day. The most popular web videos generate between 30,000 and 100,000 views per day. In this category we identified 10 very popular channels that have more than 50,000 daily views. The number of views a channel generates is the sum of its video views. Therefore, channels that publish more videos than other ones also have a higher channel view count. There are 9 extremely popular science videos with more than 100,000 views per day. Only two of them reach a number of views above the half million mark.

These videos are: *The Slow Mo Guys* ("Airbag Deploying in Slow Mo", views per day: 697,334 as of 12.3.2014) and *Vsauce* ("What Is The Resolution Of The Eye?", views per day: 528,565 as of 11.3.2014). It should be noted that both videos were uploaded less than a week before the day of data collection.

The results suggest that the most popular science videos are not always the most complex or profound ones. The productions of *The Slow Mo Guys*, for instance, are easy to understand and clearly structured: high-speed camera experiments with an entertaining presentation. *Vsauce*, as another example, is conceived as an electrifying monologue with didactic footage and seemingly trivial themes that most people take for granted such as "What is yellow?" or "Why do we kiss?"

## Description of Typologies: Design, Narrative Strategies, Genres

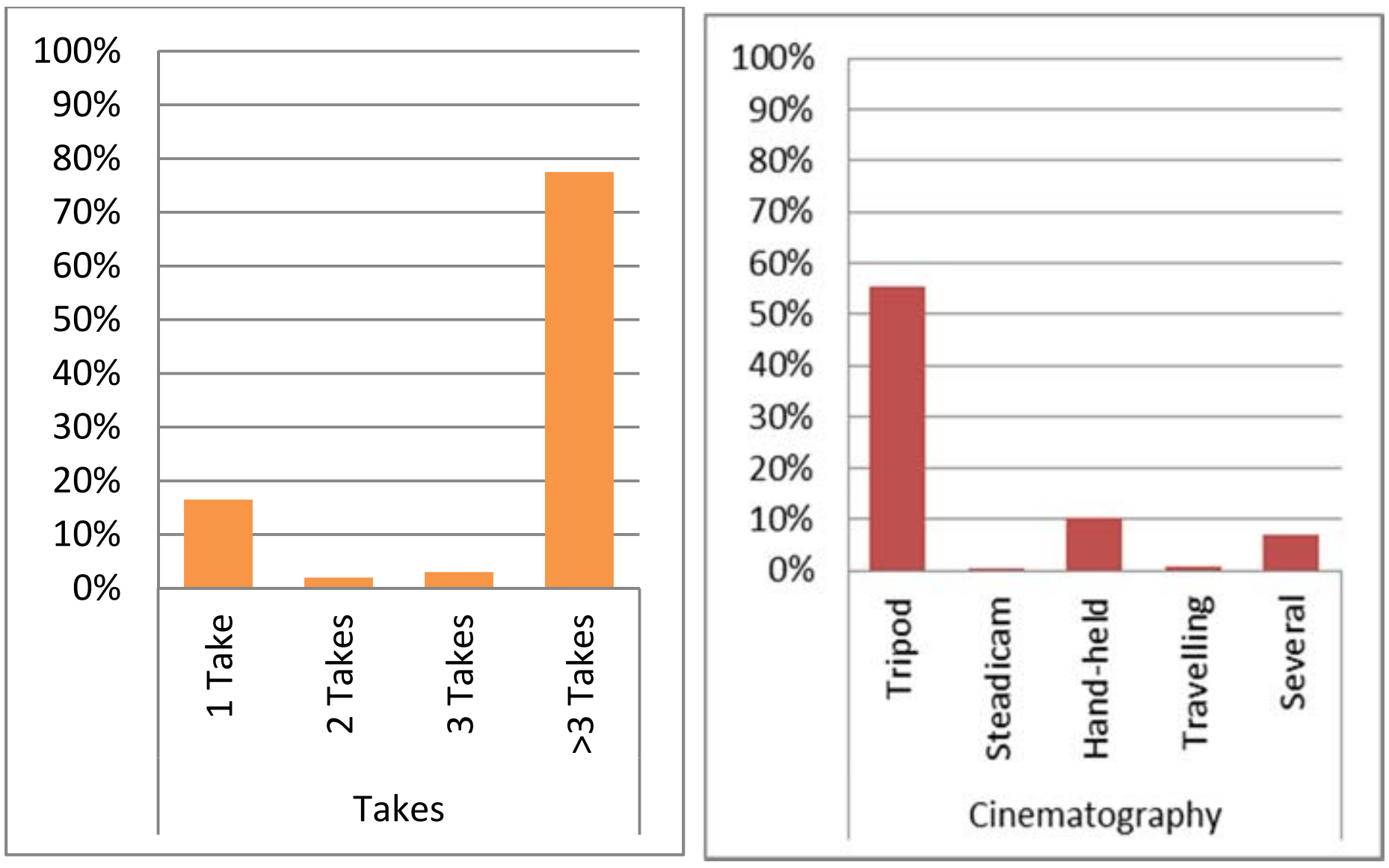

**Figures 2 and 3:** Number of takes for the production of one video, and cinematographic technique with regard to video camera supporting systems.

### Design: Montage

The amount and kind of cinematic techniques used in a video production can signify the level of professionalization with which a video was created. Examining the visual and narrative strategies deployed can therefore give a good picture of the social environment in which the work was made.

In film, the term "montage" describes the ways filmed material can be put into a coherent, final work. A take is a single, continuous video record (shot). Films and videos usually consist of numerous takes linked together, without gaps, through montage. The use of 3 or more takes can be interpreted as the result of the artist's or director's effort to construct filmic reality through montage. The use of one long take or uninterrupted shot can imply some kind of dramaturgical complexity, but since films with long uninterrupted shots are very rare and the video material we examined does not include elaborated long takes, we have to assume that web videos consisting of one take indicate a plain kind of montage. In the following section we will also discuss special effects (FX), an umbrella term for effects added after the video shoot in the post-production phase. Having a post-production phase (including but not limited to montage and FX) is also a sign of sophisticated production, as this requires a special set and level of skills.

Most of the scientific web videos have a deliberate and complex montage: over 70 % of them were produced using more than three takes (see **Figure 2**). On the other hand, there are interesting examples of scientific web videos that make do with a plan sequence, although in this case other methods like fast motion or time lapse are applied in order to match the narration for the sake of well-paced entertainment.

### Design: Cinematography

As for the cinematographic methods, we see that the most extended technique implies the use of a tripod for stable video recording (55%). Nevertheless, many of the most popular science video channels also use hand-held camera aesthetics for the production of their videos (10%), or combine multiple techniques. Brady Haran (e.g. *Sixty Symbols*), Derek Muller (e.g. *Veritasium*), and Destin Sandlin (the maker of *Smarter Every Day*), who all have extremely popular video channels, belong to the group of hand-held producers.

More elaborate cinematographic techniques such as the use of travelling and steady-cam sequences or other techniques that are usually needed for the production of scenic films are very rare at this stage of development of the scientific web video. These methods of stabilizing the video image are commonplace with professional video outlets such as TV production or cinematic movies but require sophisticated devices and specialized knowledge. A traveling camera means a camera mounted on a movable vehicle (a crane, car, lorry, or dolly system). The Steadicam system (initially developed by cameraman Garret Brown in 1975 and modified for Stanley Kubrick's "The Shining") is a camera stabilizing system where the camera is mounted on the cameraman, allowing swift movement without creating a shaky image. For simplicity, that definition includes gimbal and similar systems in our survey.

### Design: Shots

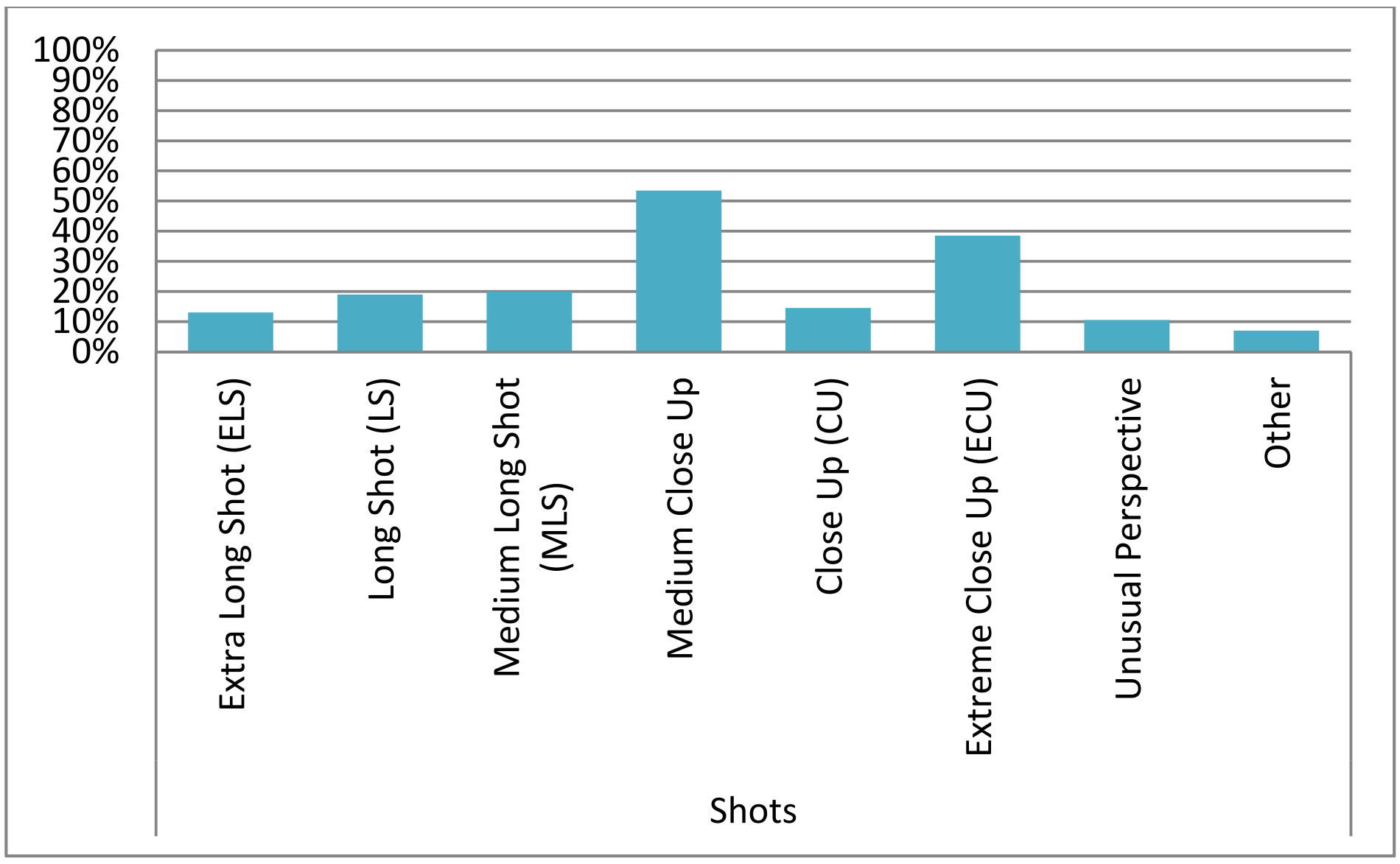

**Figure 4:** Amount of shots needed for production.

While professional stabilizing equipment beyond tripods are hardly ever used, the results revealed that the amount and variety of shots resembled the distribution of shots being used in professional documentaries, where medium close-up (MCU) and extreme close-up (ECU) shots also are the favorite ones for telling a story in pictures. Interesting is the use of close-ups for portraying people (14%) and the use of unusual perspectives in a significant number of productions. This implies, on the one hand, the importance of personal touch for this kind of video production in order to communicate scientific facts to the audience. On the other hand, the unusual perspective denotes the experimental potential that is evolving regarding this new medium on YouTube.

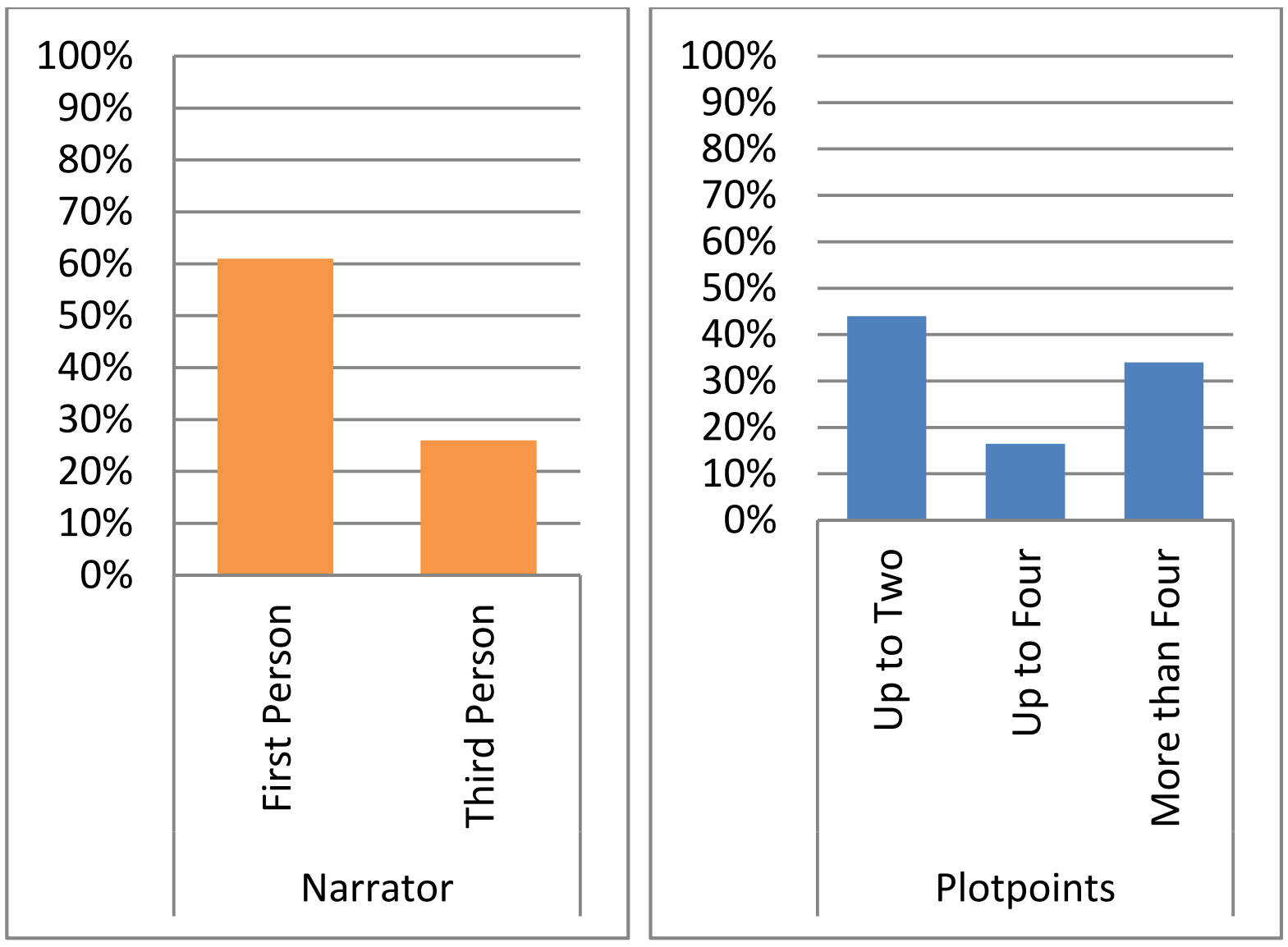

**Figure 5:** The use of narrator and the number of plot points for storytelling.

## Narrative Strategies

Most of the surveyed scientific web videos (over 60%) used a narration model in the first person, in line with the broad assumption that YouTubers seek to establish a personal connection with the audience wherein the narrator directly addresses viewers. Nevertheless, about 26 % of the YouTube video channels use a third-person narration model. Here, we have to take into account that more than one third of these productions involve animation, so that the use of a narrator as voice-over does not correspond with the structure of typical television documentary. The main motivations for the use of third-person narration seem to be its entertainment value as well as directors' attempts at innovation and originality.

We analyzed plot points to better understand the narrative strategies in scientific web videos. Plot, as a general term for a storyline, can consist of one or multiple plot points. These describe cause-and-effect turns in the narration (such as a climax, turn, rising or falling action, etc.). Generally speaking, the more plot points a storyline contains, the more complex it is.

There are many videos that use complex storytelling structures with more than 2 plot points for the development of a "scientific story" (17% with between 2 and 4 plot points, and over 33% with more than 4 plot points). Web videos that only need 2 plot points or dramatic forward movements for their "screenplay" are usually explanatory videos that consist of one question and a more or less complex structured answer to it. There can be more than one sub-plot in the answer, such as secondary explanations that lead to the end result. Therefore, even in this case there is a kind of complexity in the "screenplay".

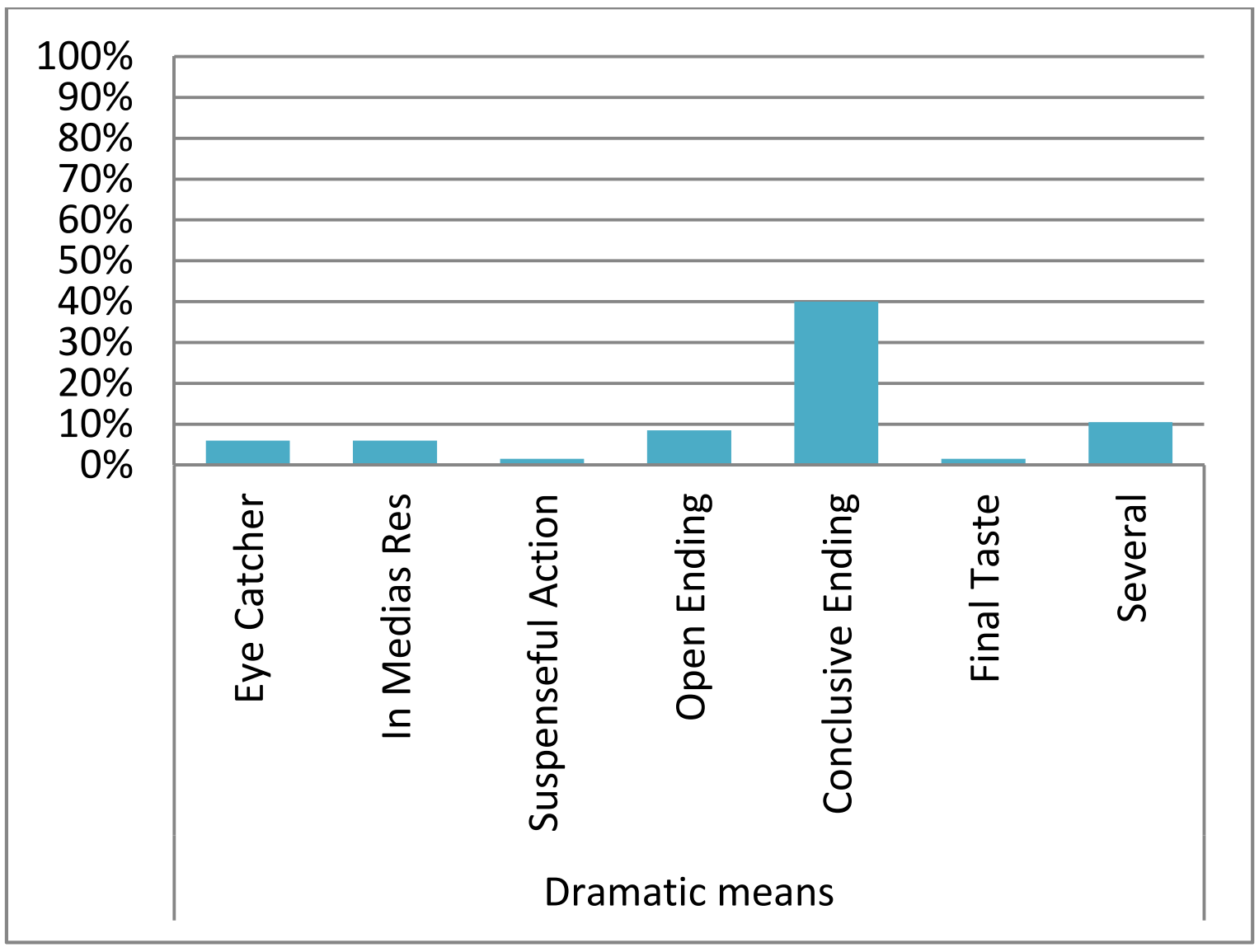

**Figure 6:** Dramatic means being used in scientific web videos.

The following dramatic means were analyzed: eye catcher, describing a beginning sequence that immediately tries to get the attention of the viewer; in medias res, which is when a video begins directly in the middle of a narration; suspenseful action, meaning special dramatic happenings; conclusive ending, i.e., a plot point that concludes the content of the video; final taste, which shows a possible outlook or positive notion at the very end of a video; or several of the abovementioned means.

If we take a closer look at the dramatic methods being used, we find a great variety of dramatic means, such as the use of an eye catcher at the beginning of a video, in medias res beginnings, or conclusive endings. Most of the dramatic energy of science web videos seems to focus on the climax at the end, which in many cases is also the answer to the formulated questions. It is remarkable that almost all scientific web videos made by Brady Haran or Derek Muller begin with an eye catcher scene that is followed by the title of the video. Some other very popular YouTubers such as the producers of *Vsauce*, *Smarter every Day*, *Sixty Symbols*, or *Veritasium* use dramatic elements to create suspenseful action, but this method is not very common among other online video educators. These YouTubers—especially those who produce moderated labor films (e.g. *The Spangler Effect*), live experiments and scientific demonstrations (e.g. *depfisicayquimica*), astronomical observations (e.g. *TheBadAstronomer*), and optical illusions (e.g. *Brussup*)—appear to be interested in narrating entertaining but very straightforward explanations of scientific facts. It can be speculated that the dramatic means used in the analyzed videos are due to the specific viewing practices on YouTube. The audience's presumed short attention span and the almost endless offerings of other videos demand different techniques from content creators.[11] Many of these creators rely on their effusive personalities, addressing their viewers as fans and dialogue partners.

The use of final taste scenes is rare. A final taste scene is an additional sequence, usually at the very end of a scenic film that provides dramatically irrelevant information in order to leave the audience with a good feeling in case the climax or ending of the film was not very positive. Suspenseful action and final taste scenes usually go together. The same directors tend to use both mechanisms in combination. One probable explanation for the often-observed absence of final taste scenes could be the short duration of most evaluated videos. Another one could be that most of the analyzed web videos don't tackle sensitive social topics.

[11] As described in the Organizational Structure of YouTube, the platform focuses on videos shorter than ten minutes. Although this formal restriction was removed in 2010, only 32 of the analyzed YouTube videos were longer than that, despite the fact that all were uploaded after the removal of time limits.

## Genres

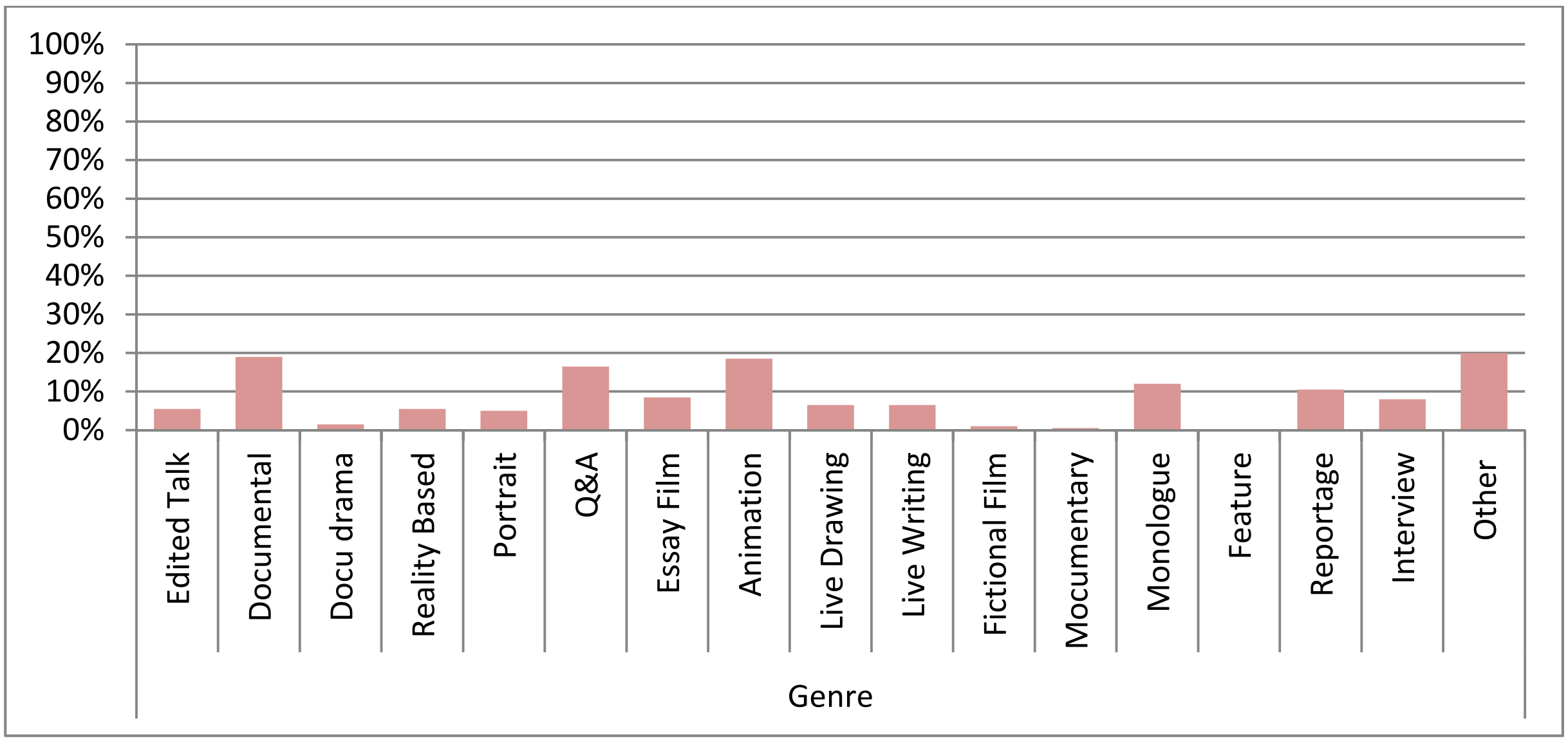

**Figure 7:** The most popular genres in the production of scientific web videos.

The variety of genres and subgenres in the production of scientific web videos for the public is manifold. As we can see in **Figure 7**, the most popular genres being used in the production of scientific web videos are the documentary and the animation movie, followed by new kind of formats that we defined as questions & answers (Q & A) and entertaining monologues on scientific topics. Classical television formats such as the reportage, the feature movie, or the interview as well as film portraits are also well represented in the investigated video corpus.

The lack of more experimental formats—more experimental than the online video itself, that is—such as fictional films, docudramas, or mockumentaries is interesting. This indicates the great focus of YouTubers on the communication of science in an entertaining but mostly very direct way. Nevertheless, there are also essay films such as the productions of PFILM (e.g. 9*4 Elements* and *Colliding Particles*), or art university projects such as the animations of the UNSWTV channel from the University of New South Wales Australia. Many of these projects focus on a new way of understanding science in a broad political and societal context. It should also be mentioned that very elaborate projects such as *94 Elements* or *Colliding Particles* and *Teslablog* are not present on YouTube, but rather rely on their own distribution platforms such as genuine websites.

The edited talk category is used rather infrequently when we look at the numbers. However, given the fact that the genre was competing with movie-oriented formats, its turnout is quite impressive. Edited talks thus seem to be a good option for the dissemination of scientific content to a broad audience: In comparison to unedited talks, these videos are mostly short (between 5 and 20 minutes) and focus on the best of the lecture, omitting slow sequences that may occur during a live talk. In addition, edited talks can even fit the lecture into a dramaturgic structure inherent to the speaker's lecture or presentation as well as to the subjective public perspective of the scene.

There is a significant amount of mixed or very specific genres (referred to as "other" in the bar diagram of **Figure 7**, i.e. 20%) that demonstrates an experimental trend within the production of scientific web videos. The most significant of them is the moderated "live experiment", as produced by YouTubers and channels such as *NurdRage Science Experiments* and the *SlowMoGuys*.

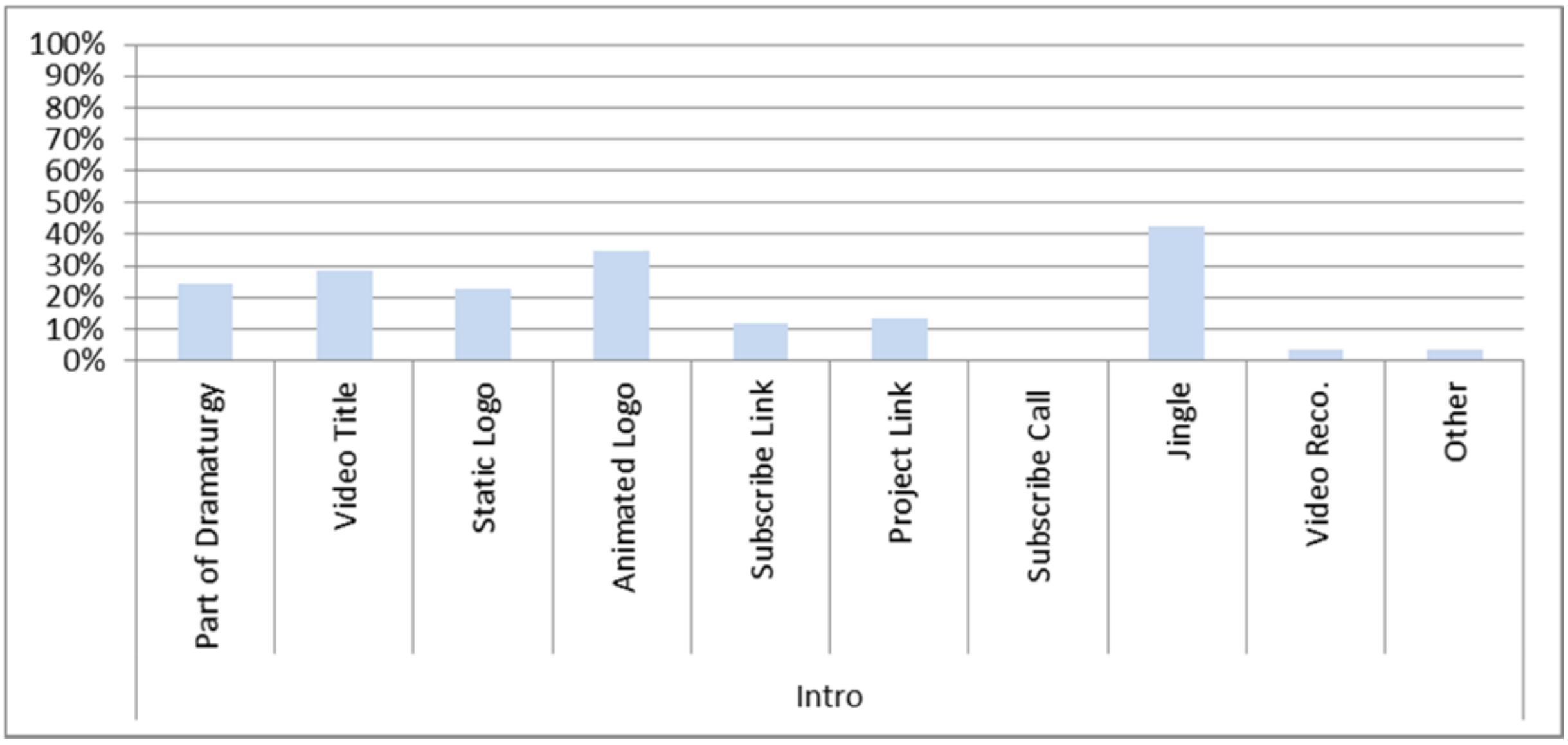

**Figure 8:** The most common elements used for the construction of intros.

The vast majority of video producers start their videos with a catchy intro. The following main characteristics were identified: Most of the videos have a characteristic, easy-to-recognize jingle tone (43%). In a large number of cases (25%) the intro sequence is already integrated in the dramaturgy of the film. As in television and cinema movies, the director does not want to waste time or lose the audience, since according to dramatic laws the first ten seconds of a web video is the amount of time you have to convince the audience of the film's entertainment value and scientific quality. In addition to this, it is very common to show the "Corporate Design" of the channel as an animated (35%) or static logo (23%) during the intro sequence. It is also interesting to note that not every web video starts with a fade-in title. Only 28% of them show the title in the intro sequence. Many channels tend to insert the title only in words or even after the intro sequence if this is part of the dramaturgy, and this then functions as an eye catcher or even as the point of attack. Subscribe links and project links are not very common in the intro section, although we did find them in 11% and 12% of the analyzed videos respectively. Invitations to subscribe are almost nonexistent.

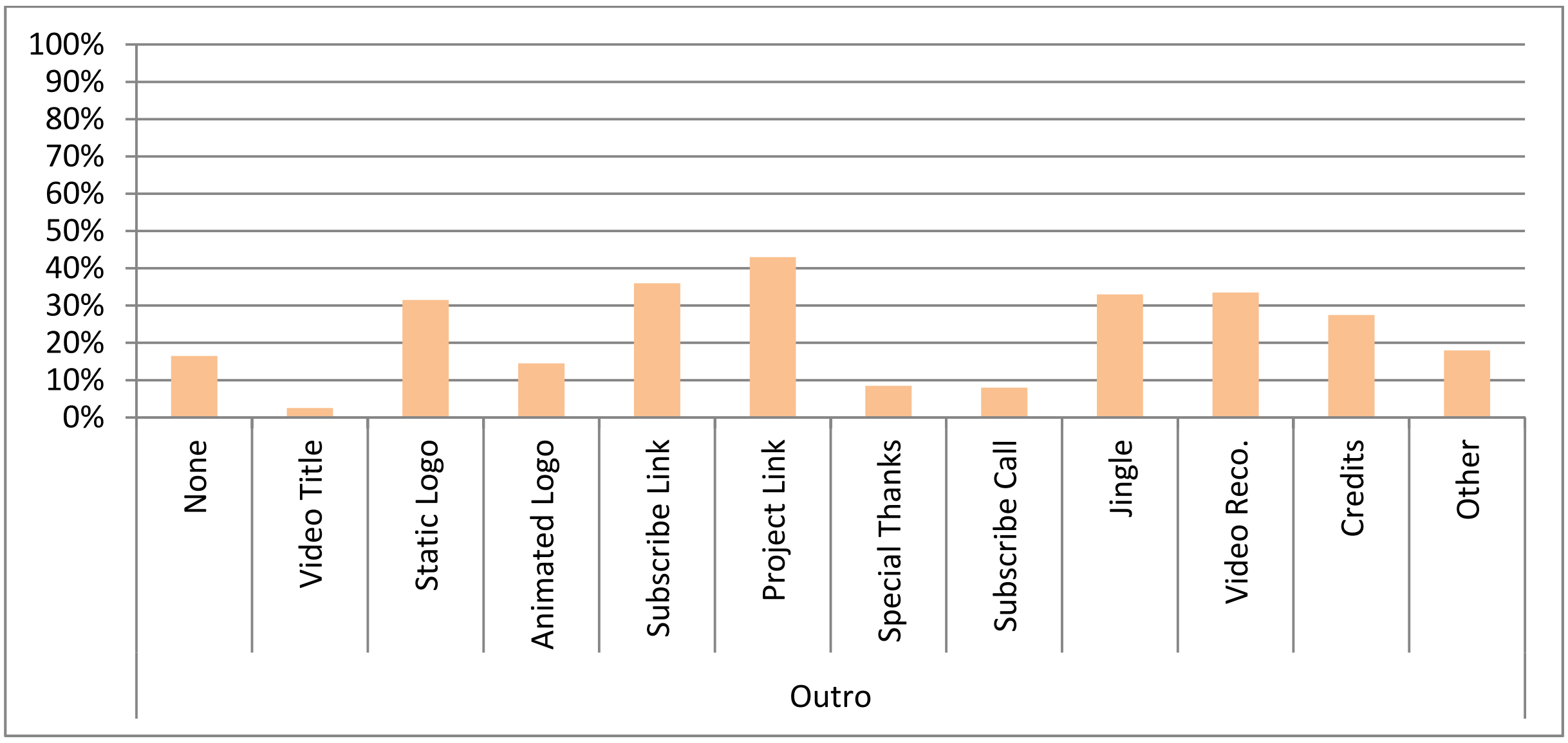

**Figure 9:** The most common elements used for the construction of outros.

The structure of outro scenes also garnered our attention. We did not find many videos with a subscribe link or an invitation to subscribe at the beginning of them, but in the outro scene this element of community building is more common: subscribe links (36%), project links (42%), and even invitations to subscribe (9%) are frequently used by YouTubers. The use of animated and static logos as well as a recognizable jingle tones is similar to what is seen in the intro sequence. Credits are not always shown. Many producers abstain from them or release the main information in the searchable description field placed underneath the video. A notable percentage of videos (18%) did not have an outro. Over 30% of producers use the outro sequence for the recommendation of other videos.

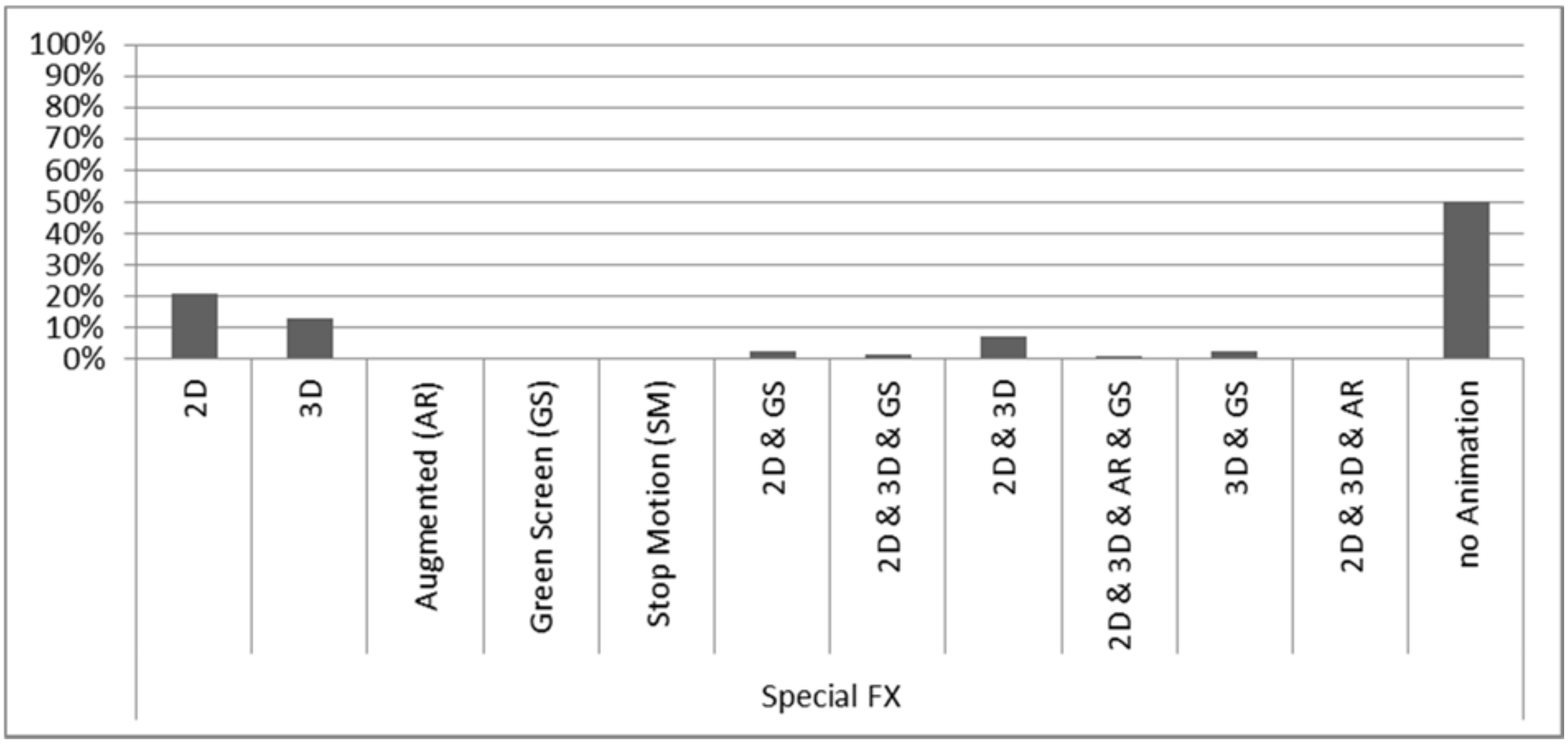

**Figure 10:** Special FX present in popular scientific videos.

Special Effects (Special FX) were used in a large number of the videos we examined. Among those are the production of 2- and 3-dimensional animations and their combination as well as other less represented film technologies such as augmented reality (AR) and green screen (GS). Nevertheless, a significant amount of the video production for educational or science communication purposes do not make use of any animation techniques. In this case, the use of simple requisites or costumes constitutes the main trends.

However, if we consider that multiple FX can be used in a single production, the percentage of videos employing Special FX could become significantly smaller. But even in this case, the results indicate that the production of scientific videos for the Internet is becoming increasingly specialized. This contradicts the general assumption that web videos, even educational ones, are being produced mostly by amateurs (Welbourne et al. 2015a).

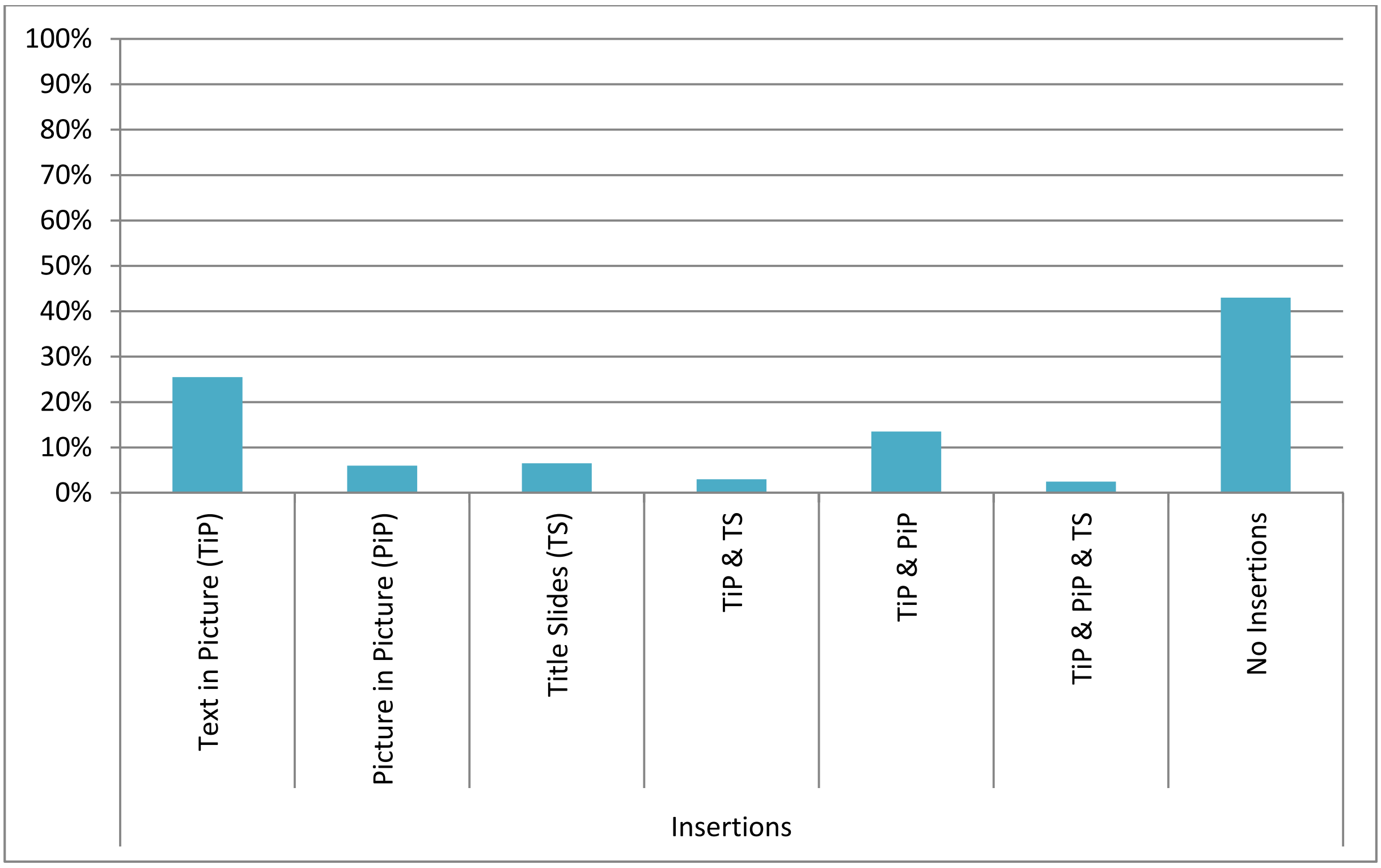

**Figure 11:** Additional FX being used in popular scientific web videos. Legend: text in picture (TiP), picture in picture (PiP), and title slides (TS).

Although 41% of the analyzed videos did not use text or picture insertions, we observed that these insertions are being increasingly used, and that they serve as didactic means for a better communication and clearer understanding of scientific facts. The use of text in picture to create an augmented reality-like composition is the most prevalent method (27%), followed by the combination of this kind of information enhancement with the use of picture in picture. As we can see in some examples such as *SmarterEveryDay*, *SciShow*, or *Vsauce*, the use of picture in picture is qualitatively manifold. While title slides and text insertions focus on additional didactic values, the use of picture in picture varies depending of the desired goal. In the outro sequences, for example, we are more likely to find picture in picture compositions as means for previews of upcoming content; while in the body sequence, this kind of additional effect has a didactic or explanatory function.

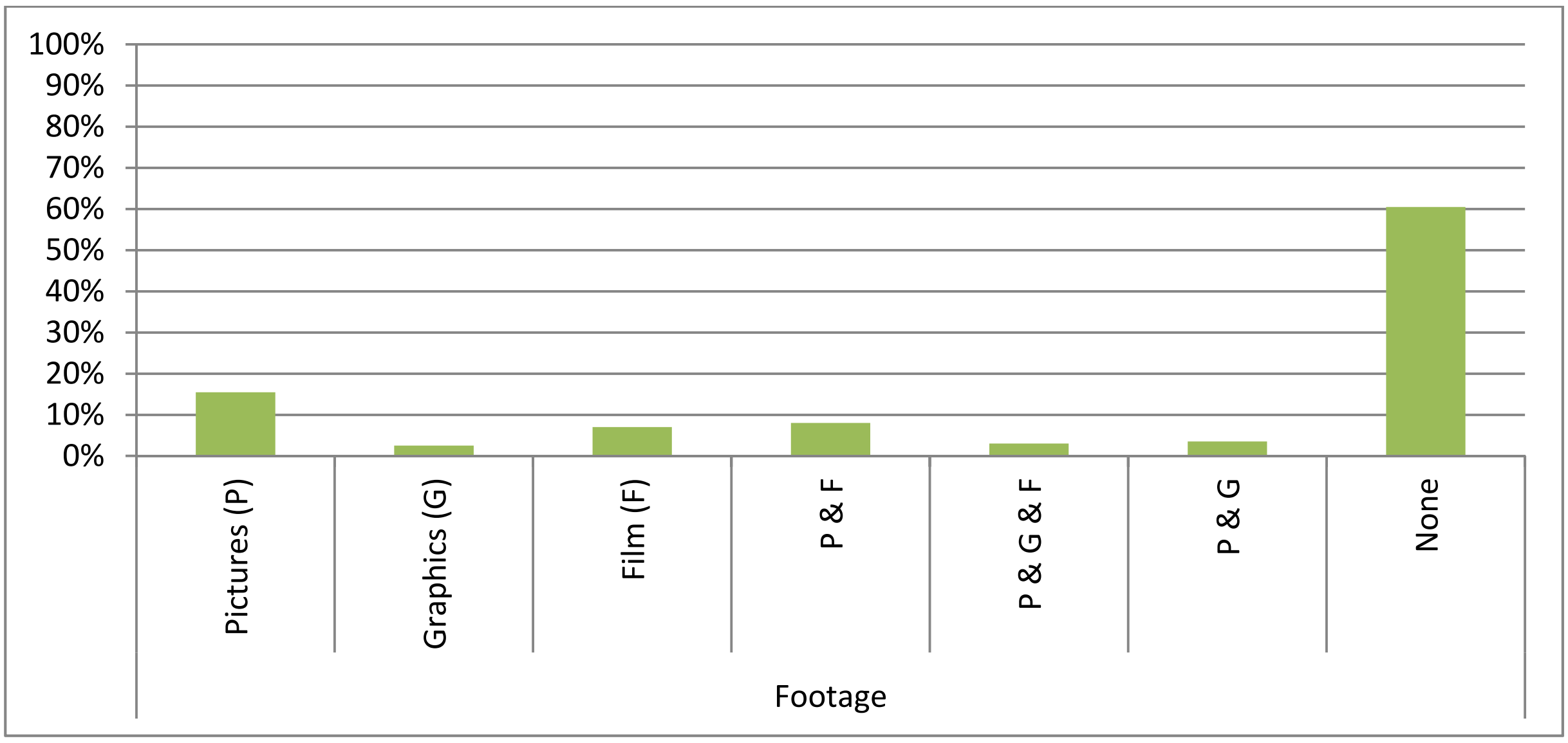

**Figure 12:** Kind of footage being used in the production of scientific web videos. Legend: pictures (P), film (F), graphics (G).

Another interesting observation is the increasing use of footage material such as pictures, old films, and graphics. This is a new phenomenon, and is particularly prevalent in video channels that deal with historical scientific facts, such as *TopZehn*, *TeslaBlogs*, *Smarter Every Day*, and *Dr. Allwissend*. There are channels that go back to free footage offered by public institutions such as international research centers or universities to quote them in their videos—mostly in the video description, but also during the video itself as a copyright disclaimer. This use of external material is actually not always safe from the perspective of possible copyright infringement—a ubiquitous topic with online video material. As an example, in German law the appropriation of third-party material for film production is only possible under the protection of the quotation law. In case of doubt, the law can decline in favor of the copyright holder, even if the presumed lawbreaker is not making commercial use of the quoted footage. Other countries such as the United States of America are less restrictive about the use of footage, despite the regulations present in the Digital Millennium Copyright Act (see, for example, Laurence Lessig's deregulation efforts and the limitations on liability relating to online material, 17 U.S. Code § 512).[12]

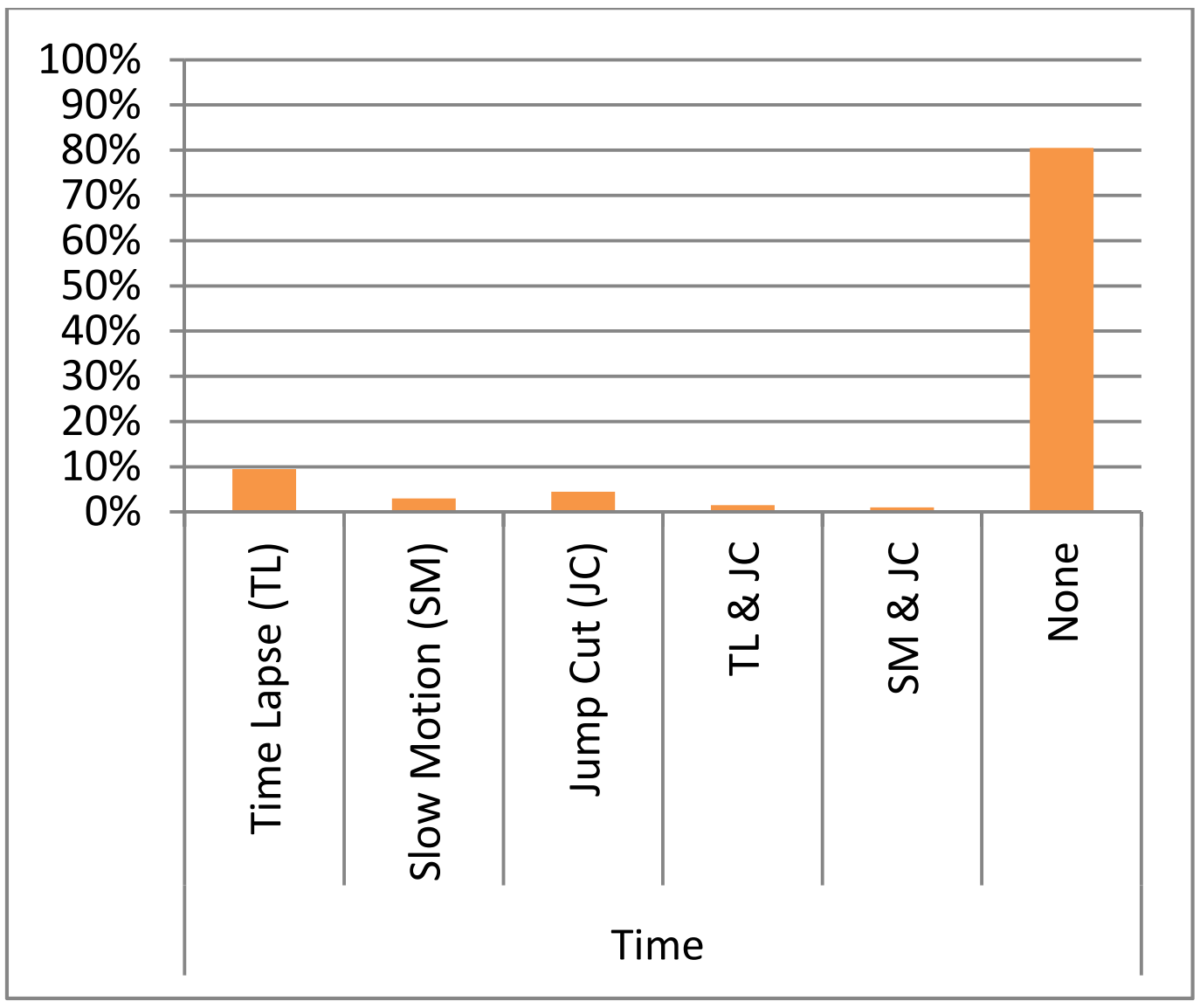

**Figure 13:** Advanced video editing techniques.

[12] http://www.law.cornell.edu/uscode/text/17/512

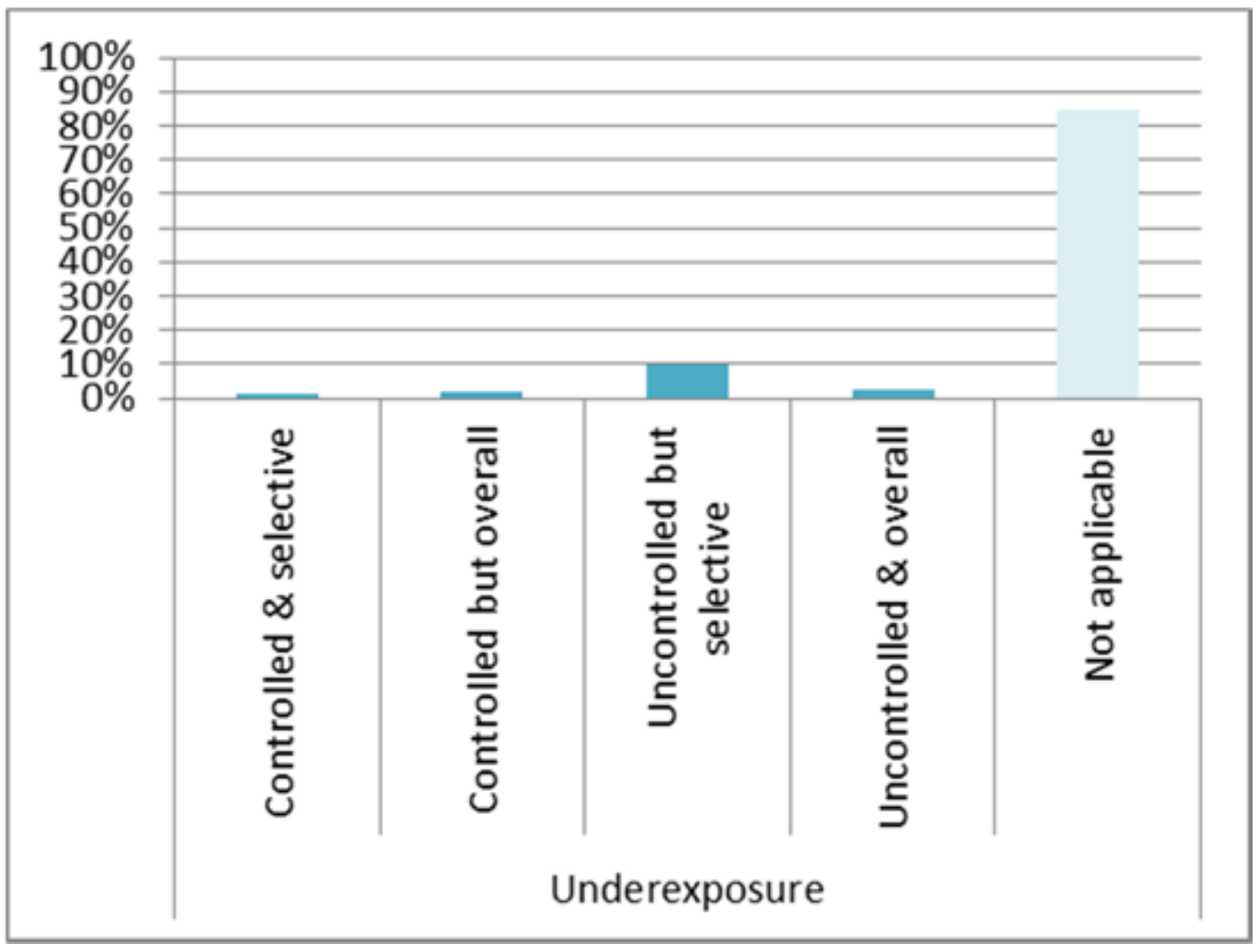

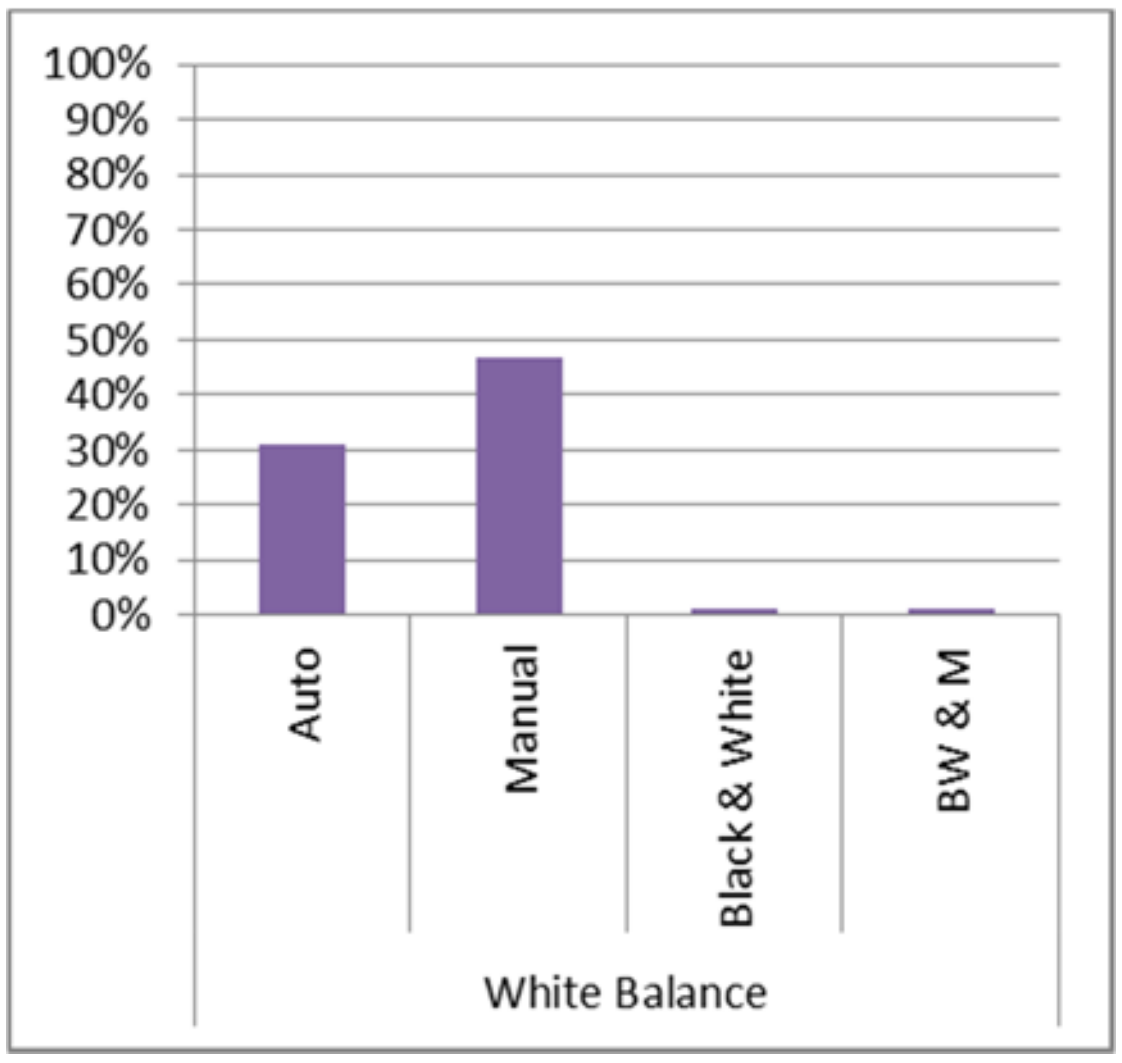

**Figures 14 and 15:** Lighting techniques and quality; type and quality of white balance.

An interesting fact is the average perception of manual white balance. Manual white balance means a manual adjustment of color toning to achieve natural color reproduction, i.e. adjusting the color sensors of a video camera to the surrounding color temperature. By doing it manually, either prior to recording or in post-production, the camera operator can generally attain the best results without changing lighting conditions.

Every change in the lighting conditions, e.g. through an unexpected change of location, must be compensated for by adjusting the color balance so as to avoid a negative impact on the image quality. For that reason, the manual white balance is a widely accepted standard among professionals. The auto white balance allows an amateur video producer to concentrate on the story without worrying about the correct representation of colors, assuming one accepts some loss in quality. Using auto white balance leaves many color discrepancies during filming, even if the location remains constant, since people or objects with different colors entering into the scene can trigger an auto white balance action even if the surrounding color context has not substantially changed. For this reason, the average use of manual white balance perceived (48%) is an indicator of the increase in the professionalism of visual productions for scientific web videos.

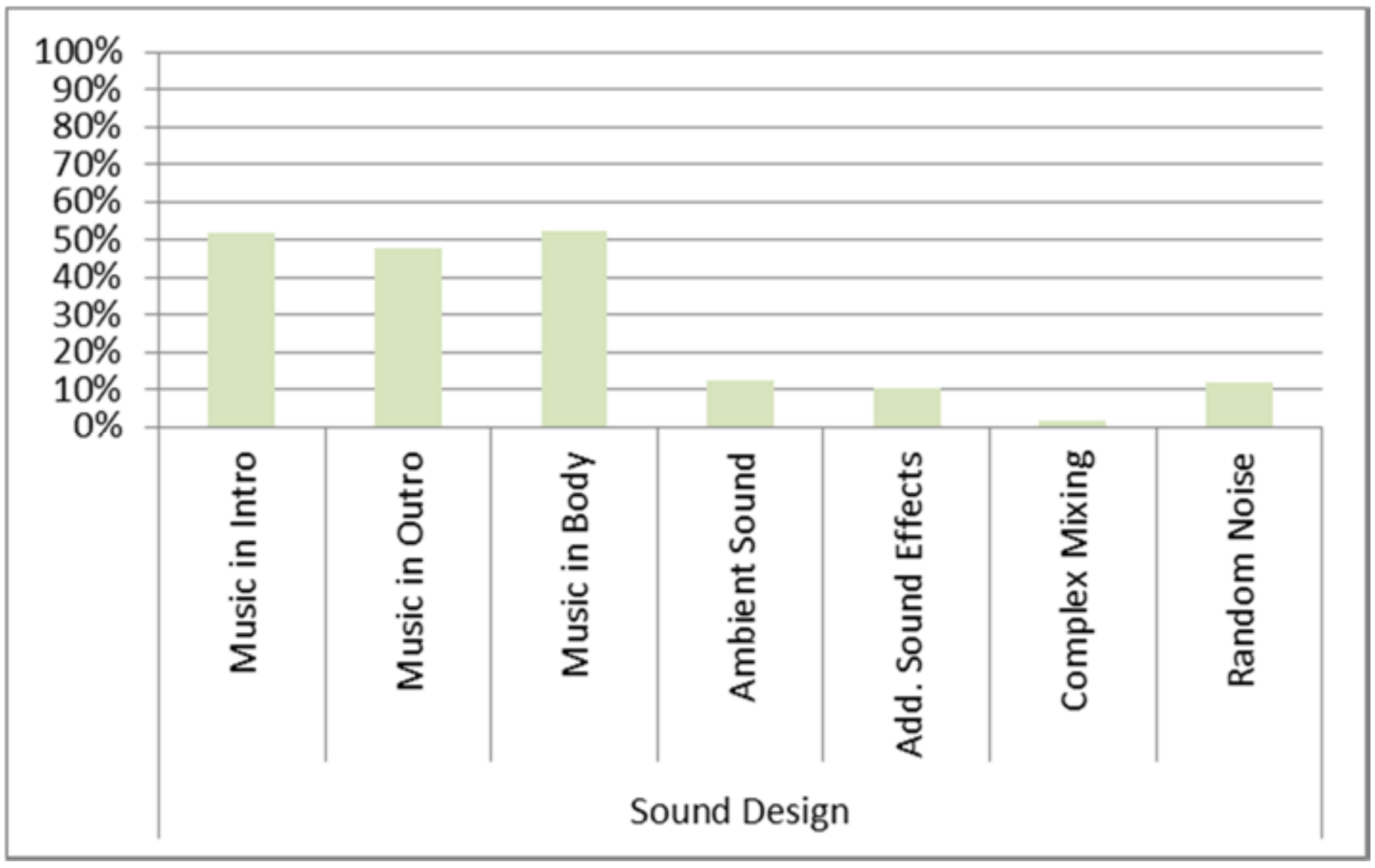

**Figure 16:** Sound design and distribution in film (indexical values).

As for sound design, more than half of the analyzed videos use some kind of accompanying music not only in the intro sequence (51%) and the outro sequence (48%) but also in the body of the film (51%). Since the acoustic level of a movie conveys a very important part of the dramaturgy of a film, additional sound effects that support storyline and suspense action as well as complex mixing of sound are indicators of a high level of professionalism in the production of scientific web videos. Furthermore, music can mask unwanted accidental background noise in a video. In this regard, random sounds (11%) and ambient sounds (12%) appear in similar proportions as additional sound effects (10%), which denotes additional production effort at the level of sound design.

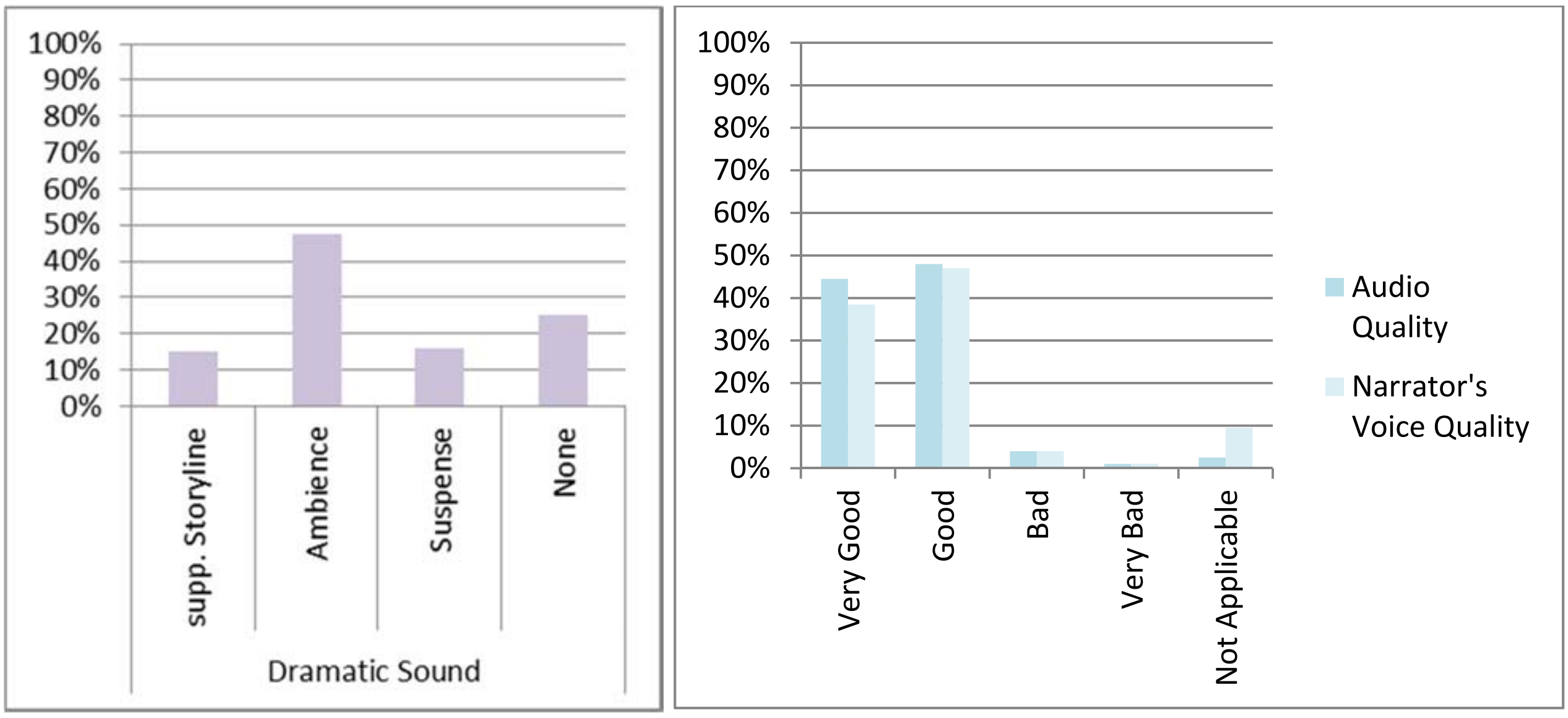

**Figures 17 and 18:** Sound design in the dramaturgic context, and audio quality.

Through the combination of music and sound effects in relation to dramatic measures, the impression of conscious or even professional sound editing becomes more noticeable: 15% of the videos use sound and music to support the storyline, and 16% of them use music and sound effects for the creation of suspenseful scenes (**Figure 17**).

In addition, the average quality of the audio and of the narrator's voice are perceived as good (49% and 48% respectively) or very good (45% and 39% respectively).

# 5. Findings

According to the commented data, we can distinguish the following characteristics and tendencies of popular science videos:

## a) English-speaking YouTube: Local multilingual vs. global English

There are very popular but not yet globally operating science video channels from non-English speaking nations: The YouTube global list is dominated by productions in the English language, with a few exceptions in Spanish (e.g. *Unicoos*) and Portuguese (e.g. *Vestibulandia*). Some of the "non-English-speaking" productions are influenced by global players from English-speaking regions. There are also original productions with the potential to become global but, for the moment, the globalization of a channel depends on translation into English and sometimes into other world languages, which often implies adjusting the content to other cultural contexts (e.g. the Spanish version of *MinutePhysics*, *MinutoDeFisica*). In this regard, if English is not the native tongue of the area, the language barrier also influences the target audiences: second-language

English speakers are usually young, well educated, and wealthy measured on a global scale (as access to video streaming is still limited to wealthier regions of the world).

**b) Variety of (sub)genres**

There is a large variety of genres and subgenres—all of them produced as short videos. The most frequent genres being used are a) documentary (e.g. *Sixty Symbols*), b) animation (e.g. *TED Ed*), and c) reportage (e.g. most university productions).

Within the main documentary and animation genres there are some subgenres that deserve a separate mention. The main subgenres are: questions & answers, which can also be found as monologues or animations; live drawing (e.g. *AsapScience*); live writing (e.g. *Khan Academy*); edited talk (e.g. *TED Talks*); portraits (e.g. *FavScientist*); or live experiments (e.g. *TheSlowMoGuys*).

**c) Moderate complexity of production**

In contrast to the popular assumption that YouTube videos are dilettante productions, we find enough evidence pointing to a certain level of professionalism, even if the use of professional or expensive recording techniques remains an exception. Nevertheless, we have discovered above-average use of some meaningful methods that point to an increase of professionalism among the producers: e.g. the manual adjustment of white balance, the use of studio lights, some recurring special FX, and the common use of tripods for stable video recording.

**d) High complexity of montage**

Most of the analyzed videos demonstrate complex montage. The number and kind of camera shots used for the elaboration of a story indicates intense post-production. Nevertheless, there are also web videos that get by with a plan sequence or less than three shots (e.g. *NurdRage*). Most of them are live experiments with either sparse moderation or none at all. In general, the variety of shots is also typical for traditional TV productions. In addition, the use of external sound devices for good voice quality and the elaboration of dramatically efficient sound design are not unusual among many producers.

**e) Storytelling experts**

The most remarkable feature and possibly the main cause of the success of web video-related science communication is the focus of YouTubers and filmmakers on storytelling. Behind every successful video is a very well-told story. Even if the main scientific topic is not a current one or is not perceived as very important for mankind, and even if the video quality is not the best, the power of an entertaining script can turn it into a viral event. This is why some channels are very successful despite certain formal production deficiencies or weaknesses such as unreliable automatic white balance (e.g. *AsapScience*), overexposure (e.g. *SmarterEveryDay* and *Sixty Symbols*), or minor sound problems.

**f) Relevance of intro and outro sequences**

Most of the online videos we analyzed use advanced intro and outro sequences. These sequences are very important for both winning new followers and keeping them informed. Many intro sequences work as a dramatic part of the movie, and most of them are conceived as eye catchers.

Recognizable elements such as animated logos and jingle tones also form part of the composition in an intro. Subscribe links and calls as well as video recommendations are common in outro sequences in order to keep growing the network of followers.

### g) The personal touch: Building an emotional network

The increasing professionalism of web video production is not so much to be found in the production techniques themselves as in the quality of movie montage and storytelling. Most of the videos are low-budget productions, and some of them may have been made by amateurs, but by amateurs with a very good knack for storytelling and *mise en scène*. Most of the short documentaries—including subcategories such as monologue, question & answer, portrait, or even interview—follow the tradition of successful TV documentaries. In that tradition, renowned scientists such as Carl Sagan or David Attenborough established a kind of personal connection with the public by explaining science directly to the camera with contagious enthusiasm. While many universities continue using standard TV reportage, the new YouTube educators keep their networks growing by addressing their audience directly and communicating via comments and video responses to questions other users may have put forward.

## 6. Conclusions

In this study, 200 science web videos from different YouTubers (university productions, professionally produced and user-generated videos) were examined to identify the main typologies and difficulty levels of video production on the Internet. We identified a wide variety of genres and subgenres, a moderate complexity of production, and a high to very high complexity of montage and storytelling that points toward growing professionalism in the production of science web videos. Most of the analyzed videos have calculated intro and outro sequences with typical elements that foster community building. The search for maximum dissemination and popularity determines the style and structure of the videos, including very short but exciting intros, a very dynamic main section, and the calculated display of subscribe links and related material at the end of every video. But the most significant aspect is that most of these YouTubers and creative web video producers are storytelling experts.

## 7. Future research: Community and Production context – toward a network analysis

The research presented provides an initial description of the state of the art in an unmanageable and ever-changing field: the science video in the world of online video production. Research on the typologies of online science videos has only just begun. The researchers' next goal is a survey of production context, types of video descriptions, and the specific YouTube recommendation cultures—e.g. which links are recommended where and why. We assume a logical and consequent interrelation among typologies and contextual aspects, and observe that the code book we have used already hints at most of this information. For the moment, we have opted for a separate discussion of the results for the sake of clarity, dividing them into identity-related characteristics on the one hand and contextual-environmental aspects on the other. Studying the production context in detail would go beyond the scope of the present publication, whose main purpose is an initial description of common science video typologies.

In our upcoming research, however, the correlation, causation, and comparison of production context with the kind of video production used will allow new insights into the nature of the video typologies already investigated. Indeed, in order to see the whole picture, information on scientific online video typologies should be enhanced by including information about the videos' production context and the recommendation culture they are a part of. In addition, a network analysis of the interrelations of the producers will also provide an insightful assessment of the remarkable diversity of producers as well as of the sheer dimensions of the phenomenon itself.

## 8. Acknowledgments

The authors would like to thank Friederike Shymura, Klarissa Niedermeier, and Thi Hoai Thuong Truong for their assistance in watching the videos and collecting a large amount of the video data, as well as Kareem James Abu-Zeid, Silke Flörchinger, Alina Marktanner, and Jens Görisch for their helpful comments and editing, including some grammatical corrections to this paper.

# 9. Appendix

**Table 1: Coding questionnaire**

*Questionnaire for data collection on science web videos for the public*

<table>
<tr><th colspan="4">Information on the Video Channel (Important: Please doublecheck all data!)</th></tr>
<tr><td colspan="4">Consecutive Number from Dataset: #</td></tr>
<tr><td colspan="4">Video Title:</td></tr>
<tr><td colspan="4">Date:</td></tr>
<tr><td colspan="4">Filled out by:</td></tr>
<tr><th colspan="4">Video: Design description</th></tr>
<tr><td>Number of actors according to gender</td><td colspan="3">Female actor(s)/speakers:<br>Male actor(s)/speakers:</td></tr>
<tr><td>Thumbnail</td><td>☐ Picture from video<br>☐ Edited picture</td><td>Thumbnail<br>(description)</td><td></td></tr>
<tr><td>Estimated age of the actor(s)</td><td>Female actor(s):<br>☐ <= 17 ☐ 18-25<br>☐ 26-35 ☐ 36-45<br>☐ 46-55 ☐ 56-65<br>☐ 66-75 ☐ 76-85 ☐ >85:<br>Male actor(s):<br>☐ <= 17 ☐ 18-25<br>☐ 26-35 ☐ 36-45<br>☐ 46-55 ☐ 56-65<br>☐ 66-75 ☐ 76-85 ☐ >85:</td><td>Location(s)<br>(multiple choice)</td><td>☐ Indoor without stage setting<br>☐ Indoor with stage setting<br>☐ Outdoor without stage setting<br>☐ Outdoor with stage setting<br>☐ Other:</td></tr>
<tr><td>Camera work<br>(multiple choice)</td><td>☐ One-take story<br>☐ Two-take story<br>☐ Three-take story<br>☐ More than three takes<br>----------------------------------<br>☐ Tripod shooting<br>☐ Steadicam<br>☐ Hand-held camera<br>☐ Travelling camera<br>☐ Other:</td><td>Shots used<br>(multiple choice)</td><td>☐ Extreme long shot<br>☐ Long shot<br>☐ Medium long shot<br>☐ Medium shot<br>☐ Medium close-up<br>☐ Close-up<br>☐ Extreme close-up<br>☐ Unusual perspective/shot<br>☐ Other:</td></tr>
<tr><td>Storyline<br>(multiple choice)</td><td>☐ First-person narrator<br>☐ Third-person narrator<br>Number of plot points:<br>☐ Up to 2 plot points<br>☐ Up to 4 plot points<br>☐ More than 4 plot points<br>☐ Eye catcher in opening<br>☐ In medias res opening<br>☐ Suspenseful action<br>☐ Open ending<br>☐ Conclusive ending<br>☐ Final taste<br>☐ Other:</td><td>Genre<br>(multiple choice)</td><td>☐ Edited talk ☐ Monologue<br>☐ Documentary ☐ Feature<br>☐ Docudrama ☐ Reportage<br>☐ Reality based ☐ Interview<br>☐ Portrait<br>☐ Questions & Answers<br>☐ Essay film<br>☐ Animation/Cartoon<br>☐ Live drawing<br>☐ Live writing<br>☐ Fictional film<br>☐ Docufiction / Mockumentary<br>☐ Other:</td></tr>
</table>

<table>
<tr>
<td>Intro<br>(multiple choice)</td>
<td>☐ No intro<br>☐ Part of dramaturgy section<br>☐ Video title<br>☐ Static logo<br>☐ Animated logo<br>☐ Subscribe link<br>☐ Project or author’s link<br>☐ Subscribe call (spoken)<br>☐ Channel jingle<br>☐ Video recommendation(s)<br>☐ Other:</td>
<td>Outro<br>(multiple choice)</td>
<td colspan="2">☐ Video title<br>☐ Static logo<br>☐ Animated logo<br>☐ Subscribe link<br>☐ Project or author’s link<br>☐ Special thanks<br>☐ Subscribe call (spoken)<br>☐ Channel jingle<br>☐ Video recommendation(s)<br>☐ Credits<br>☐ Other:</td>
</tr>
<tr>
<td>FX and Light<br>(multiple choice)</td>
<td>☐ 2D animation<br>☐ 3D animation<br>☐ Augmented reality<br>☐ Green screen<br>☐ Stop motion<br>-----------------------------<br>☐ Text-In-Picture<br>☐ Picture-In-Picture<br>☐ Title slide(s)<br>------------------------------<br>Archive material<br>☐ Pictures<br>☐ Graphics<br>☐ Video footage<br>-------------------------------<br>☐ Time lapse<br>☐ Slow motion<br>☐ Jump cut<br>------------------------------<br>HD quality: ☐ Yes ☐ No<br>☐ Other:</td>
<td>☐ Studio lights<br>☐ Available light<br>☐ Daylight<br>☐ Artificial light<br>☐ Controlled light<br>☐ Overexposure<br>☐ Underexposure<br>☐ Selective<br>☐ Overall<br>☐ Uncontrolled light<br>☐ Overexposure<br>☐ Underexposure<br>☐ Selective<br>☐ Overall<br>☐ Other:</td>
<td>White balance (WB)</td>
<td>☐ Auto WB<br>☐ Manual WB<br><br>☐ Black & white<br>☐ Other:</td>
</tr>
<tr>
<td>Sound design<br>(multiple choice)</td>
<td>☐ Music in intro<br>☐ Music in outro<br>☐ Music in body<br>☐ Ambient sound<br>☐ Additional sound effects<br>☐ Complex sound mixing<br>☐ Random noise<br>☐ Other:</td>
<td>Kind of music<br>(multiple choice)</td>
<td colspan="2">☐ Music supports storyline<br>☐ Ambience music<br>☐ Suspense music<br>☐ No music<br>☐ Other:</td>
</tr>
</table>

| | | | |
|---|---|---|---|
| **Audio quality** | ☐ good ☐ very good<br>☐ bad ☐ very bad | **Narrator's voice quality** | ☐ good ☐ very good<br>☐ bad ☐ very bad |
| **Recommended Links** | (see table below)<br># Number of Links<br>**Where? I** Intro/**B** Body/ **O** Outro/ **D** Description (Each "Where" gets its own line!)<br>**Which kind? Di** Discussion/ **Info** Information/ **FM** Film Maker Site / **Sub** Subscribe / **Do** Donate/ **TW** Twitter/ **FB** Facebook/ **OV** Other Videos/ **VTC** Video Time Code/ **Other?** Describe! | | |

| Where? | # | Which kind? |
|---|---|---|
| | | |
| | | |
| | | |
| | | |
| | | |

| | | | |
|---|---|---|---|
| **Video: Production context** | | | |
| **Production context** | ☐ Individual Production<br>☐ Profit<br>☐ Non-Profit<br>  ☐ Museum<br>  ☐ University<br>  ☐ NGO<br>  ☐ Other:<br>☐ Not found | **Production firm** *(Name and link)* | |
| **Sponsoring** *(Names and links)* | | | |
| **Other** | | | |

*Thank you for your cooperation!*

**Table 1: Full List of 200 analyzed videos and channels**

| Channel | Video title |
|---|---|
| asapSCIENCE | Which Came First – The Chicken or the Egg? |
| asapSCIENCE | The Scientific Power of Teamwork |
| Khan Academy | Basic Addition |
| Khan Academy | Subtracting complex numbers |
| MinutePhysics | Immovable Object vs. Unstoppable Force - Which wins? |
| Minute Physics | How modern Light Bulbs Work |
| NurdRage | Coke Can in Liquid Nitrogen |
| NurdRage | Make an Iron Heart |
| Scientific American Space Lab | How to Enter the Space Lab Competition |
| Scientific American Space Lab | Behind the Scenes |
| SciShow | The Truth About Gingers |
| SciShow | Trouble in Bed: When Sleep Turns Against Us |
| SpanglerScienceTV | Magic Sand - Sand that is Always Dry! |
| SpanglerScienceTV | Lava Lamp - Cool Science Experiment |
| Sixty Symbols | Putting your Hand on the LHC |
| Sixty Symbols | The Sound of Atoms Bonding |
| Smarter Every Day | How Houdini Dies (in Slow Motion) |
| Smarter Every Day | Cold Hard Science. The Physics of Skating on Ice |
| SpaceRip | Earth 100 Million Years from Now |
| SpaceRip | Water Planet |
| TED Ed | Questions no one knows the answer to |
| TED Ed | Why do we cry? The three types of tears |
| TED Talks | Tony Robbins: Why we do what we do |
| TED Talks | Mary Lou Jepsen: Could future devices read images from our brain? |
| Veritasium | World's Roundest Object |
| Veritasium | Can you solve this? |
| Vi Hart | Hexaflexagons |
| Vi Hart | Cookie Shapes |
| SickScience | Dry Ice Boo Bubbles |
| SickScience | Power of Bleach |
| Acchiappamente | 2x05 - Disney e Coca-cola ti controllano? [Messaggi subliminali - Psicologia] |
| Acchiappamente | #Acchiappamente - Stress buono o cattivo? |
| Alberto Lori | Libertà di cambiare (psicologia quantistica) |
| Alberto Lori | Il pensiero focalizzato (psico quantistica) |
| ANUchannel | Richard Dawkins and Lawrence Krauss: Something from Nothing |
| ANUchannel | A Conversation with Andrew Macintyre |
| BozemanScience | A Tour of the Cell |
| BozemanScience | The Brain |
| Canal Educatif à la demande (CED) | Simulation d'entretien de recrutement . |
| Canal Educatif à la demande (CED) | L'art en Question 08: Carpaccio - Le Jeune Chevalier |
| Deep Sky Videos | Messier Objects |
| Deep Sky Videos | Inside an Opening Telescope |
| La Educación Prohibida | LEP - Archivos Abiertos #01 - Carlos González . |
| La Educación Prohibida | LEP - Archivos Abiertos #11 - Antonio Solórzano . |
| Geração de Valor | GV #08 - Atitude + Preparo = Sucesso . |
| Geração de Valor | Geração de Valor - Casos de sucesso . |
| GeroMovie | DNA- Replication Biologie |
| GeroMovie | Winkelarten |
| ImbaTorben | Todesmilch Titten |
| ImbaTorben | Die 5 A's - so bekommst du du jede Frau |
| matemarika86 | Non funziona la funzione!!! - Studio di funzioni e dominio |
| matemarika86 | VIDEO INTERATTIVO Caccia al tesoro: Alla ricerca della X perduta - Equazioni di primo grado intere |
| Mental Floss | 50 Common Misconceptions |
| Mental Floss | 27 Amazing Facts about Comics |
| NASA JPL Videos | Mars Science Laboratory Curiosity Rover Animation |
| NASA JPL Videos | What's Up for March 2014? |
| The Slow Mo Guys | Giant 6ft Water Balloon |
| The Slow Mo Guys | Airbag Deploying in Slow Mo |
| NewScientist | Spray-on Clothing |
| NewScientist | Cyborg Drummer creates Unique Beats |
| Nucleus Medical | Birth: McRoberts Maneuver |
| Nucleus Medical | Nucleus Custom Medical Animation Process |
| PBS IdeaChannel | Are Bronies Changing the Definition of Masculinity? |
| PBS IdeaChannel | Does Twitch Plays Pokemon Give You Hope for Humanity? |
| RMIT University | How hydrogen engines work |
| RMIT University | Australia-India Research Centre für Automation Software Engenieering |
| Scientific American | Your Brain in Love and Lust |
| Scientific American | Is Our Universe a Hologram? |
| Storm | Amazing Upward Lightning! |
| Storm | Extreme Dust Storm Takes Over Phoenix, Arizona 2011 |
| T Log | Sitchin & Anunnaki 1/6 |
| T Log | Vuoto Quantomeccanico |
| foodskey | Being mean to broccoli |
| foodskey | Phytoplasmas in Plants |
| Unicoos | Matriz inversa, traspuesta y adjunta 2ºBACHI unicoos matematicas |
| Unicoos | BILLION = BILLON?? unicoos nosvemosenclase Facebook compra Whatsapp |
| Unisciel | Faire implose une canette |
| Unisciel | Unisciel Select: numero 55 |
| Universcience | FIV mode d'emploi |
| Universcience | Herbier #7 - on a une belle série de citrons |
| UNSW TV | How to survive beach rip current |
| UNSW TV | Why winds explain earth's surface warming slowdown |
| CrashCourse | The Agricultural Revolution |
| CrashCourse | Fate, Family, and Oedipus Rex |
| Computerphile | The Problem with Time & Timezones |
| Computerphile | EXTRA BITS - Installing Ubuntu Permanently |
| MinuteEarth | Where Did Earth's Water Come From? |
| MinuteEarth | Are any Animals Truly Monogamous? |
| Numberphile | Why do YouTube views freeze at 301? |
| Numberphile | Brussels Sprouts |
| PeriodicVideos | Gold Bullion Vault - Periodic Table of Videos |
| PeriodicVideos | The world's greatest autograph book |
| Vsauce | What if Everyone JUMPED at once? |
| Vsauce | What is the Resolution of the Eye? |
| The SpanglerEffect | Getting Ready for Guiness World Record Season 01 Ep.01 |
| The SpanglerEffect | Flying Toilet Paper Season 02 Ep.19 |
| CGPGrey | The Difference between the UK, GB and England |
| CGPGrey | The Law You Won't Be Told |
| Vlogbrothers | Giraffe Love and Other Questions ANSWERED |
| Vlogbrothers | Is the American Dream Real? |
| Quantum Fracture | ¿Qué es la Ciencia? |
| Quantum Fracture | Uno de los Principios Más Importantes del Universo |
| MinutoDeFísica | Errores comunes en física |
| MinutoDeFísica | $E=mc^2$ está Incompleta |
| Ever Salazar | Calculando Areas |
| Ever Salazar | (not so) Cold Fun: Qué hacer chuando no está tan frio afuera |
| The Spirit Science | Spirit Science 1 ~ Thoughts |
| The Spirit Science | Spirit Science 22 (part 4) ~ Source Energy |
| ScienceBob | Science Bob's Crazy Foam Experiment |
| ScienceBob | Exploding Pumpkins on Jimmy Kimmel Live |
| ouLEarn | Shakespeare: Original Pronunciation |
| ouLEarn | Maryam Bibi - Unlikely Leaders (2/5) |
| Euronews Knowledge | World's smallest atomic clock |
| Euronews Knowledge | Can Earthquakes Bring Life? Do You Know? |
| Naked Scientists | How does DNA fingerprinting work? Naked Science Scrapbook |
| Naked Scientists | Main Alu II re-uploaded |
| Educatina | Síntesis de proteínas - Biología - Educatina |
| Educatina | Patrones de medición - Física - Educatina |
| FavScientist | Gregor Mendel - My Favourite Scientest |
| FavScientist | Ignaz Semmelweis - My Favourite Scientist |
| Depfisicayquimica | Agua que no cae / The water doesn't fall down |
| Depfisicayquimica | Cubo que no se derrama II |
| Brusspup | Amazing Anamorphic Illusions |
| Brusspup | Cool Rolling Illusion Toy! How to |
| Jörn Loviscach | 22.6.1 Stetigkeit |
| Jörn Loviscach | P3 Datumsdifferenz in Tagen mit Embedded Controller |
| ChemExperimentalist | How to make sulfuric acid |
| ChemExperimentalist | Make Calcium Hydroxide - $Ca (OH)^2$ from Plaster of Paris |
| Abenteuer Wissen | Star Trek: Wie funktioniert Impuls- und Warpantrieb |
| Abenteuer Wissen | Timescapes: Learning to Fly - die Welt im Zeitraffer |
| Welt der Wissenschaft | Animaterie und Relativität |
| Welt der Wissenschaft | Wie einzigartig ist der Mensch? |
| Welt der Physik | Was ist ein schwarzes Loch? |
| Welt der Physik | Monsterwellen im Labor |
| WissensMagazin | Der tiefste Blick ins All |
| WissensMagazin | $E=mc^2$ - Die Äquivalenz von Masse und Energie |
| TheSimpleMaths | Exponentialfunktion und Logarithmus |
| TheSimpleMaths | Gebrochenrationale Funktionen - Nullstellen, Definitionsbereich… |
| Fisica Total | Física Total - Aula 07 - vetor - Vetores e operações vetoriais |
| Fisica Total | ENEM em AÇÃO - Física #01 (principais habilidades cobradas na prova de…) |
| Canal Ciência e Ficção | Jurassic Park parte 1/2 - É possível clonar dinossauro? - Ciência e Ficção |
| Canal Ciência e Ficção | Star Wars Episódio VII - O Que Esperar? - Ciência e Ficção |
| Manual do Mundo | Congele água em 1 seg - o segredo |
| Manual do Mundo | Revelação da Mágica dos ladrões de galinha (mágica fácil revelada) |
| Quirkology | 10 More Amazing Science Stunts (3) |
| Quirkology | The Tube of Mystery |
| NASA eClips | Real World: Space Shuttle Thermal Protection System |
| NASA eClips | Real World: Comet Quest |
| It's Okay To Be Smart | The Science of Snowflakes |
| It's Okay To Be Smart | How The Elements Got Their Names |
| Bill Nye the Science Guy | Atoms |
| Bill Nye the Science Guy | Climate |
| TheBadAstronomy | Snow that doens't melt! Is it a government conspiracy? |
| TheBadAstronomy | Glory from an Airplane window |
| EEVblog | World's Most Expensive Hard Drive Teardown |
| EEVblog | Voltech PM300 Power Analyser Teardown |
| NobelPrize | Interview with 1994 Laureate in Economics John Nash |
| NobelPrize | Ben Bernanke speaks about the 2011 Laureates in Economic Science |
| SmithsonianChannel | Titanoboa: Monster Snake: Titanoboa vs. T-Rex |
| SmithsonianChannel | Secrets of the Third Reich: This Video Exposes Hitler's Secret Illness |
| ScienceChannel | Jumping Jack Ants vs. Huntsman Spider/ Monster Bug Wars |
| ScienceChannel | The Time Scientists Thought They Saw The Real Death Star |
| Stevebd1 | Nuclear Fusion |
| Stevebd1 | A Sudden Multiplication of Planets |
| techNyouvids | Critical Thinking Part 1: A Valuable Argument |
| techNyouvids | Nanotechnology Part 2: Nanoproperties |
| Explainity | Euro-Krise leicht erklärt |
| Explainity | Korruptionsbekämpfung leicht erklärt |
| Doktor Allwissend | Warum alle Apple lieben |
| Doktor Allwissend | Warum wir Vorurteile brauchen |
| Getty Museum | The Mummification Process |
| Getty Museum | Ansel Adams: Visualizing a photograph |
| Bite Szi-zed | Caffeine |
| Bite Szi-zed | Reindeer Eyes |
| Nature Video | Lego Antikythera Mechanism |
| Nature Video | The Beginning of Everything |
| Brainscoop | Where My Ladies At? |
| Brainscoop | The Two-Faced Calf, Pt. I |
| Wahre Verbrechen.Wahre Stories | Doktor Allwissends ABC der Kriminalität #Y wie Yakuza |
| Wahre Verbrechen.Wahre Stories | Fritz Haarmann - SERIAL KILLERS #WV.WS |
| Northwestern NewsCenter | The Incredible Robot Fish |
| Northwestern NewsCenter | James Agaard, A Lifetime Wildcat |
| DLRde | DLR crawler robot meets Justin and Hand-Arm-System |
| DLRde | How Philae got ist name |
| TU Muenchen | Typisch TUM |
| TU Muenchen | TUM Ambassador Professor Patrick Dewilde about the importance of networks for scientific thinking |
| European Space Agency, ESA | First-ever live 3D video stream from space |
| European Space Agency, ESA | Sentinel-1A rides into space on a Soyuz |
| Kurzgesagt | Fracking explained: opportunity or danger |
| Kurzgesagt | Engineering & Curiosity |
| Hybrid Libraryian | World's 10 Most Mysterious Pictures Ever Taken |
| Hybrid Libraryian | Earth's 10 Most Important Events in History |
| Cambridge University | Memories of old awake |
| Cambridge University | Putting our House in Order: William Kent's Designs for the Houses of Parliament |
| Colliding Particles | Codename Eurostar |
| Colliding Particles | The Dice |
| 94 elements | Copper-Acid and Dust |
| 94 elements | Osmium-Fingerprints |
| Teslablog | Tesla y los superpoderes II: Invisibilidad |
| Teslablog | Tesla y la máquina del tiempo |